\documentclass[10pt,twocolumn]{revtex4-1}

\usepackage{amsmath, amsfonts, amssymb}     %
\usepackage{graphicx}                       %
\usepackage{xfrac}                          %
\usepackage{grffile}                        %
\usepackage{color,colortbl}

\usepackage[colorlinks, citecolor=blue, urlcolor=red, linkcolor=blue, hypertexnames=false]{hyperref}

\def\arraystretch{1.5}

\newcommand{\tsr}[1] {\mathbf{#1}}
\newcommand{\vtr}[1] {\mathbf{#1}}

\newcommand{\xhat}{\hat{\mathbf{x}}}
\newcommand{\yhat}{\hat{\mathbf{y}}}
\newcommand{\zhat}{\hat{\mathbf{z}}}

\newcommand{\gdot}[0] {\dot{\gamma}}
\newcommand{\gdotb}[0] {\bar{\dot{\gamma}}}
\newcommand{\degree}{\ensuremath{^\circ}}

\newcommand{\lC}[0]{\ensuremath{\ell_C}}
\newcommand{\lm}[0]{\ensuremath{\ell_\mu}}

\newcommand{\vecv}[1]{\mathbf{{#1}}}
\newcommand{\tens}[1]{\mathbf{{#1}}}

\newcommand{\bi}{\begin{itemize}}
\newcommand{\ei}{\end{itemize}}

\newcommand{\be}{\begin{equation}}
\newcommand{\ee}{\end{equation}}
\newcommand{\bea}{\begin{eqnarray}}
\newcommand{\eea}{\end{eqnarray}}

\newcommand{\etas}{\eta_{\rm s}}
\newcommand{\etaa}{\eta_{\rm a}}

\begin{document}
\title{Edge fracture instability in sheared complex fluids: onset criterion and possible mitigation strategy}
\author{Ewan J. Hemingway and Suzanne M. Fielding}
\affiliation{Department of Physics, Durham University, Science Laboratories,
  South Road, Durham DH1 3LE, UK}
\date{\today}
\begin{abstract}
  We perform a detailed theoretical study of the edge fracture instability, which commonly destabilises the fluid-air interface during strong shear flows of entangled polymeric fluids, leading to unreliable rheological measurements.  By means of direct nonlinear simulations, we map out phase diagrams showing the degree of edge fracture in the plane of the surface tension of the fluid-air interface and the imposed shear rate, within the Giesekus and Johnson-Segalman models, for different values of the nonlinear constitutive parameters that determine the dependencies on shear rate of the shear and normal stresses. The threshold for the onset of edge fracture is shown to be relatively robust against variations in the wetting angle where the fluid-air interface meets the hard walls of the flow cell, whereas the nonlinear dynamics  depend strongly on wetting angle.  We perform a linear stability calculation to derive an exact analytical expression for
  the onset of edge fracture, expressed in terms of the shear-rate derivative of
  the second normal stress difference, the shear-rate
  derivative of the shear stress (sometimes called the tangent viscosity), the jump in shear stress across the
  interface between the fluid and the outside air,
  the surface tension of that interface, and the rheometer gap size. (The shear stress to which we refer is $\sigma_{xy}$ with $\xhat$ the flow direction and $\yhat$ the flow gradient direction. The interface normal is in the vorticity direction $\zhat$.) Full agreement between our analytical calculation and nonlinear simulations is demonstrated.   We also elucidate in detail the mechanism of edge fracture, and finally suggest a new way in which it might be mitigated in experimental practice. We also suggest that, in containing the second normal stress difference, our criterion for the onset of edge fracture may be used as a means to determine that quantity experimentally. Some of the results in this paper were first announced in an earlier letter~\cite{hemingway2017edge}. The present manuscript provides additional simulation results, calculational details of the linear stability analysis, and more detailed discussion of the significance and limitations of our findings.
\end{abstract}
\date{\today}
\maketitle

\section{Introduction}

\begin{figure}[!t]
  \centering
  \includegraphics[width=0.45\textwidth]{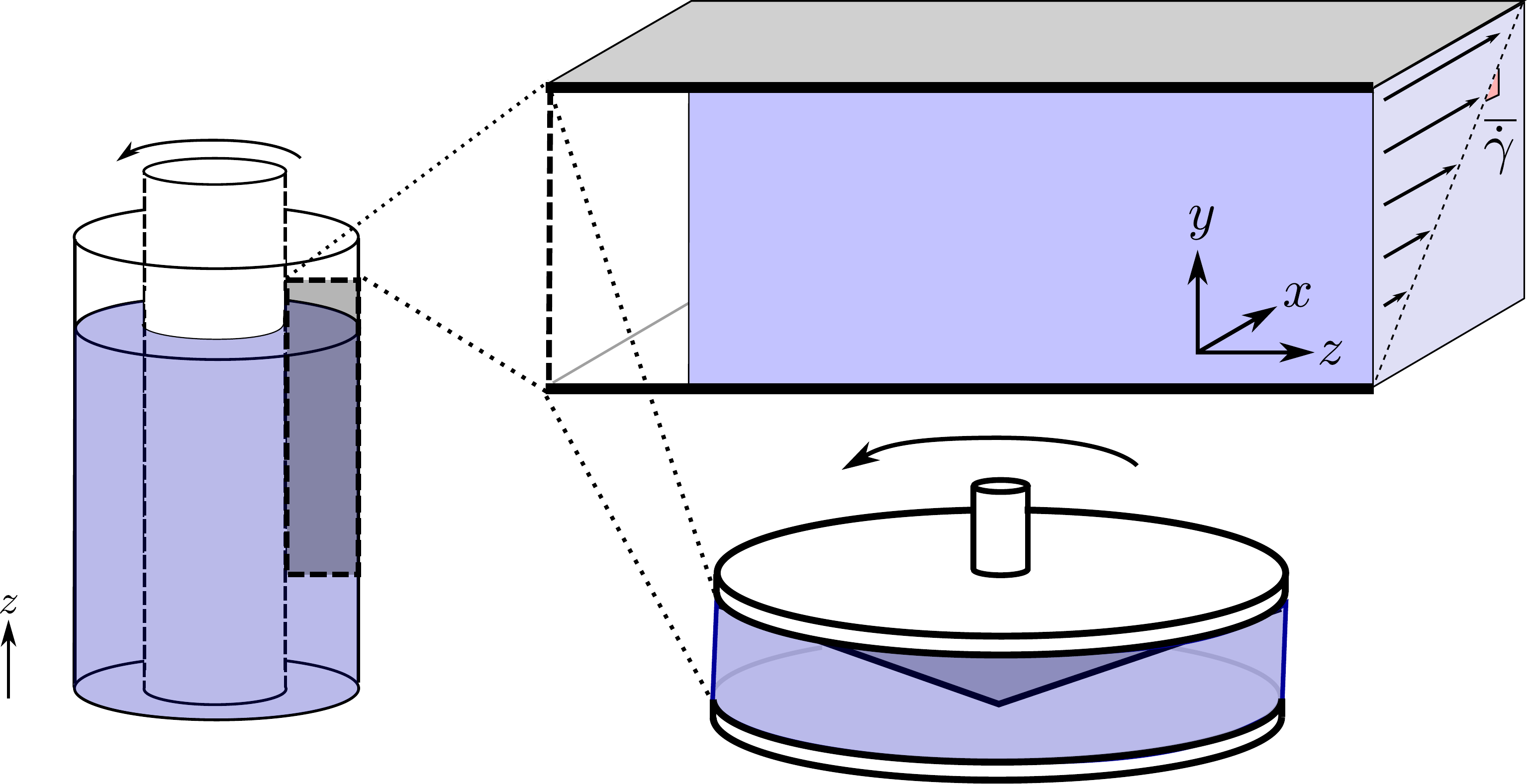}
  \caption{Schematic showing cylindrical Couette flow (left) and cone-plate flow (bottom). Also shown is the way in which both of these flows are approximated in our simulations by planar Couette flow (top right). In each case the fluid (blue) has an interface with the outside air (white).}
  \label{fig:schematic}
\end{figure}

Measurements of a fluid's shear rheology are commonly performed in a torsional flow device, using either a cone-and-plate or cylindrical Couette  flow cell.  The former comprises a cone rotating relative to a stationary plate (or vice versa). The latter comprises two coaxial cylinders,  with the inner cylinder rotating relative to the outer (or vice versa). See the schematic  Fig.~\ref{fig:schematic}. In each case, the speed of rotation normalised by the gap size (or angle) sets the imposed shear rate, $\gdot$. The resulting torque gives the shear stress response, $\sigma$.  In a state of steady flow, the shear stress $\sigma$ as a function of shear rate gives the flow curve, $\sigma(\gdot)$, which is key to characterising a material's rheological response. Other common tests of a fluid's shear rheology include oscillatory shear,  shear startup, step stress, and stress relaxation tests.

Rheological measurements performed in the regime of linear viscoelastic response are typically well controlled and highly reproducible. In contrast, the measurement of stronger, nonlinear flows are much more challenging. This is particularly true in highly viscoelastic materials such as entangled polymer melts and concentrated polymer solutions, as well as in concentrated suspensions. Attempts to measure the steady state flow curve or transient flow behaviour at high strain rate (or stress) are often beset by flow instabilities that can lead to unreliable data. Many such instabilities depend not only on the bulk rheology of the material in question, but also on the geometry of the flow device used, including the boundary conditions where the fluid sample meets the hard walls of the flow cell, and/or the outside air.

In both of the flow devices sketched in Fig.~\ref{fig:schematic}, the fluid (shown in blue) has an interface with the outside air (shown in white). In  entangled polymers or concentrated suspensions, this
free surface is highly susceptible to  destabilising when the material is strongly sheared, particularly in a cone-and-plate or plate-plate device. Above a critical imposed shear rate, the sample edge will deform into a more complicated edge profile, forming an indentation that invades the fluid bulk.  In violent cases, some portion of the sample can even be ejected from the measurement region. 

This phenomenon, which is known as `edge-fracture', renders accurate
measurements of strong flows extremely difficult. Indeed, to quote Snijkers~\cite{snijkers2011cone}, ``the effects of edge fracture on rheological measurements performed with standard rotational rheometers using cone-and-plate or plate-plate geometries are catastrophic and render the torque (in a strain-controlled rheometer) or rotation speed (in a stress-controlled rheometer) as measures of bulk rheological properties of the fluid, virtually useless".
Jensen described the phenomenon as  ``the limiting factor in
rotational rheometry''~\cite{jensen2008measurements}.  Detailed
experimental studies of edge fracture can be found in
Refs.~\cite{lee1992does,inn2005effect,sui2007instability,schweizer2008departure,mattes2008analysis,jensen2008measurements,dai2013viscometric}. Anecdotal reports further pervade the  literature.  In recent decades, strategies to mitigate edge fracture have been developed, involving specialised guard-ring~\cite{mall2002normal} or cone-partitioned-plate devices~\cite{schweizer2003comparing,schweizer2004nonlinear,snijkers2011cone,meissner1989measuring,schweizer2013cone,costanzo2018measuring}.

 Despite the crucial importance of edge fracture to experimental shear rheology, the phenomenon has remained poorly understood theoretically, until recently. In insightful early works, Tanner and
coworkers~\cite{tanner1983shear,keentok1999edge} predicted that edge fracture should arise
for a critical magnitude  of the second normal
stress difference, $|N_2(\gdot)|>\Gamma/R$, where $\Gamma$ is the surface tension of the fluid-air interface and $R$ is some pre-assumed
geometrical lengthscale. 
Taken as a scaling argument, this prediction shows remarkable
insight. Indeed, a careful experimental study later confirmed the key role of $N_2$ (rather than the first normal stress difference, $N_1$) in driving edge fracture~\cite{lee1992does,keentok1999edge}. However, Tanner's prediction will prove only partly consistent with our findings below. In particular, it fails to incorporate the role of the shear stress, alongside the second normal stress, in driving edge fracture.

In a recent Letter~\cite{hemingway2017edge}, we performed a detailed theoretical study of edge fracture, combining linear stability analysis with full nonlinear simulations. Outcomes of this work that we hope will be useful to the experimental rheology community include: (i) a new criterion for the onset of edge fracture that now recognises the importance, alongside $N_2$, of the slope of the flow curve of shear stress as a function of shear rate, and the jump in shear stress between the fluid sample and outside air; (ii) a full mechanistic understanding of the edge fracture instability; and (iii) a new suggestion for how edge fracture might potentially be mitigated in experimental practice.

The present paper aims to provide a fuller discussion of the results
originally announced in Ref.~\cite{hemingway2017edge}. In particular,
we set out here for the first time the details of the analytic linear
stability calculation by which we arrived at the new onset
criterion. We also provide new simulation results, beyond those in
Ref.~\cite{hemingway2017edge}, delineating more fully the dependence
of the onset threshold on the nonlinear parameters of the two
constitutive models considered (which set the scaling of the shear and
normal stresses with strain rate), and demonstrating the role of the
wetting angle in the mode of edge fracture. For definiteness we
cast the discussion mostly in the language of entangled
polymers. However, we note that the constitutive models used here are
not restricted to such materials alone, and our results may indeed apply more generally to other classes of fluids. It would be particularly interesting in future work to consider the edge fracture instability in non-Brownian suspensions and in viscoelastic yield stress materials.

Besides edge fracture, another
important instability that routinely confounds attempts to measure strong flows of complex fluids is that of wall slip~\cite{hatzikiriakos2015slip,hatzikiriakos2012wall}: the layer of fluid immediately adjacent to the hard wall of the flow cell shows an apparent slip relative to the wall itself. We shall ignore wall slip in what follows, suppressing the phenomenon upfront by assuming conditions of no-slip. This is potentially a major shortcoming of our approach, which should be addressed in future studies. 

The paper is structured as follows. In Sec.~\ref{sec:geometry} we introduce the flow geometry to be studied. Sec.~\ref{sec:models} sets out  the models and methods that we shall use. The results of our nonlinear simulations are discussed in Sec.~\ref{sec:nonlinear}, followed in Sec.~\ref{sec:linear} by a linear stability analysis to understand the phase boundary for the onset of edge fracture, as obtained in our nonlinear simulations. This linear analysis also permits a detailed understanding of the mechanism of instability. In Sec.~\ref{sec:mitigation}, we suggest a way in which our findings indicate a possible practical route to mitigating edge fracture experimentally. Finally, Sec.~\ref{sec:conclusion} provides conclusions and perspectives for future work.

\section{Flow geometry}
\label{sec:geometry}

As noted above, measurements of a fluid's shear rheology are often performed in a rotational flow device, commonly using either a cone-and-plate or cylindrical Couette flow cell, as shown schematically in Fig.~\ref{fig:schematic}. For the former device, we assume a small cone angle and large cell radius. In this limit, the curvature of the streamlines becomes negligible and the flow is well approximated by a planar slab of fluid sheared between flat plates, as sketched in Fig.~\ref{fig:schematic} (top right). (In experimental practice, however, we note that very small cone angles are difficult to obtain because of the limited stiffness of rheometer frames; while large samples are forbidden by the limited normal force capacity of the transducers and the volume of sample available.) For the latter device, we consider the limit $\delta R/R\ll 1$, in which the gap between the cylinders, $\delta R$, is small compared with the inner cylinder radius, $R$.  In this limit, the flow is again well approximated by a planar slab. 

\begin{figure}[!t]
  \includegraphics[width=0.45\textwidth]{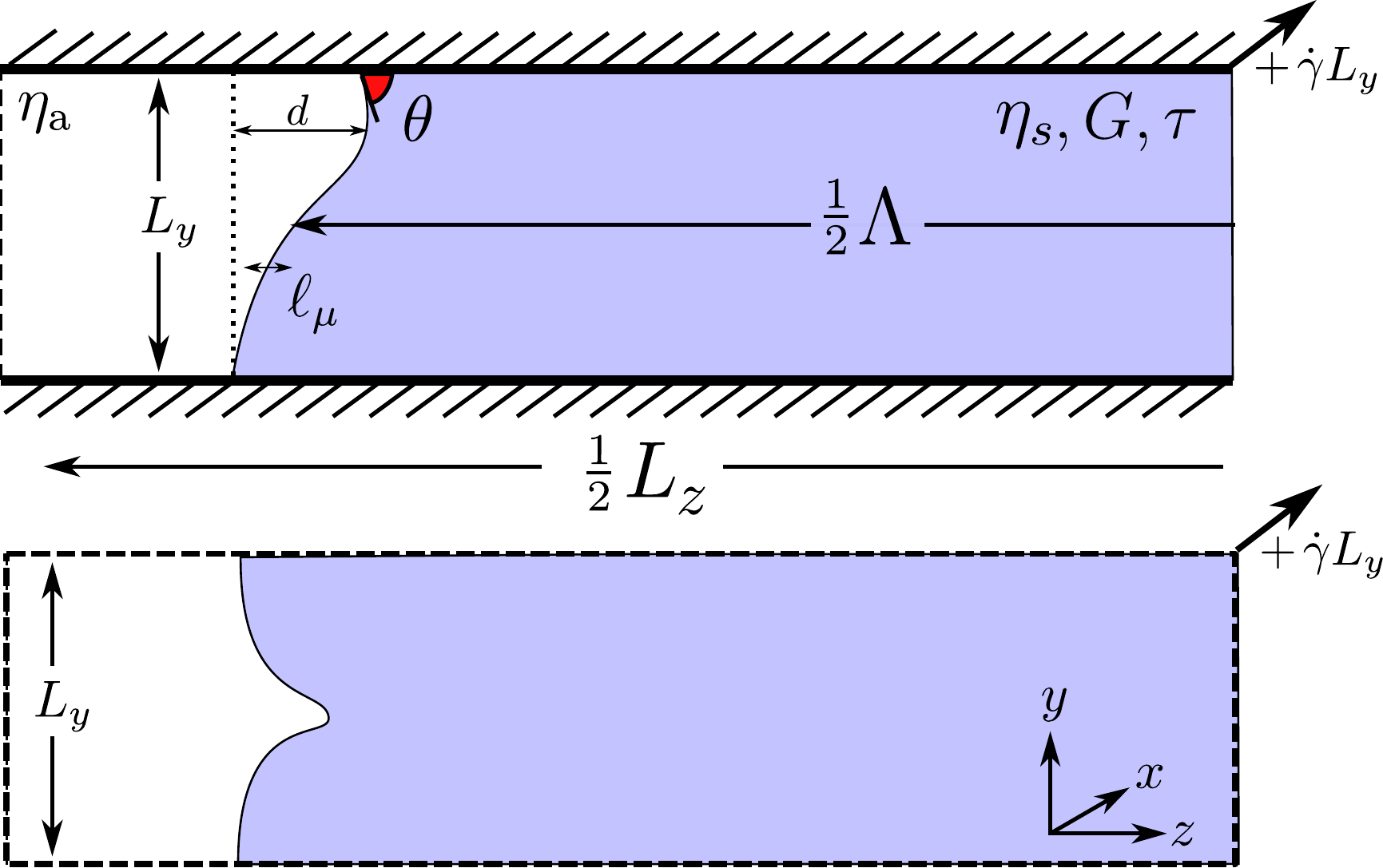}
  \caption{Sketch of flow geometries simulated. Top: planar shear between hard flat plates at $y=\pm L_y/2$. Bottom: planar shear with Lees-Edwards periodic boundary conditions. The symbols denoting material properties and geometry dimensions are defined in the main text.}
  \label{fig:geometry}
\end{figure}

Throughout what follows, therefore, we shall consider a planar slab of fluid in contact with the outside air, as sketched again in Fig.~\ref{fig:geometry}.
The sample is sheared at some rate $\gdot$ by moving the top boundary at some speed $\gdot L_y$ in the flow direction $\xhat$, into the page. We consider always positive shear rates, $\gdot>0$. (For most fluids, the second normal stress $N_2(\gdot)\sim-\gdot^2$ at low shear rates. For negative flow rates we would need to replace $d|N_2|/d\gdot$, which appears several times below, by $d|N_2|/d|\gdot|$.) 

We denote the flow-gradient direction by $\yhat$, shown vertically in Fig.~\ref{fig:geometry}. The edges of the sample in the vorticity direction $\zhat$ (horizontal) are in contact with the air. Only the left half of the box is shown in
Fig.~\ref{fig:geometry}: an equivalent interface exists in the right half. We assume that the flow remains translationally invariant in the flow direction, $\xhat$, setting $\partial_x$ of all quantities equal to zero, and performing two-dimensional (2D) calculations in the gradient-vorticity plane
$y-z$. (All simulation snapshots  below will therefore show only this $y-z$ plane.) The velocity vector and stress tensor are themselves however 3D objects.  This assumption of only 2D variations should be checked in future, fully 3D studies.

The sample length in the vorticity direction $z$ at the cell mid-height $y=0$ (before the fluid is sheared) will be denoted
$\Lambda$. The simulation box has length $L_z$ with periodic boundary
conditions in the vorticity direction. In the flow-gradient direction, $y$, we consider two different kinds of boundary condition.  The first models  hard walls at $y=\pm L_y/2$, with boundary conditions of no slip or
permeation at these walls. The second considers the simplified theoretical geometry of
Lees-Edwards biperiodic shear. In that case, all quantities repeat
periodically across shear-mapped points on the boundaries of box
copies stacked in the $y$ direction, but with adjacent copies moving
relative to each other at velocity $\gdot L_y \hat{\vtr{x}}$. The threshold for the onset of edge fracture obtained in our numerical simulations will
prove to be in excellent agreement between these two cases. Our reason for invoking the simplified biperiodic
geometry is that it will allow progress in analytical calculations, which would
otherwise be prohibitively complicated.

\section{Models}
\label{sec:models}

\subsection{Force balance and incompressibility}

We assume that inertia is negligible, and work in the creeping flow limit of zero Reynolds number. In this limit, the condition of force balance requires the total stress in any element of fluid (or air), $\tens{T}(\vecv{r},t)$, to obey:
\be
\nabla.\tens{T}=0.
\ee
As usual, $\vecv{r}$ denotes position and $t$ time.

Inside the fluid, we assume $\tens{T}$ to comprise an isotropic contribution described by a pressure $p(\vecv{r},t)$, a
Newtonian contribution of viscosity $\etas$, and a
viscoelastic contribution $\tens{\Sigma}(\vecv{r},t)$ stemming from the complex fluid
microstructure (polymer chains, etc.). The condition of force balance inside the fluid then reads:
\be
\etas\nabla^2\vecv{v}+\nabla.\tens{\Sigma}-\nabla p=0,
\ee
in which 
$\vecv{v}(\vecv{r},t)$ is the fluid velocity.

The air outside the fluid lacks any viscoelastic component, $\tens\Sigma=0$, and has a lower viscosity, $\etaa\ll\etas$, giving the force balance condition:
\be
\etaa\nabla^2\vecv{v}-\nabla p=0
\ee
The pressure field $p(\vecv{r},t)$ is determined by assuming the flow to be everywhere incompressible, with the flow velocity
$\vecv{v}(\vecv{r},t)$ obeying 
\be
\nabla.\vecv{v}=0.
\ee
Note that in computing everywhere in the fluid and air the flow rate, our simulation naturally captures the changes shear rate that will be present in the vicinity of any edge-fracturing disturbances in the fluid-air interface.

\subsection{Constitutive models}

The dynamics of the viscoelastic stress $\tens{\Sigma}$ in flow is
determined by a viscoelastic constitutive equation. In what follows,
we shall study two constitutive models that are widely used across the
rheological literature.  Our aim in studying two different models is
to establish the degree to which any predictions concerning the onset
of edge fracture are generic across constitutive models, or whether
they instead depend on model details. Importantly, indeed, we shall
show that the predictions of these two models for edge fracture depend
on their respective parameters $a$ or $\alpha$ (defined below) only
via the appearance of those quantities in the shear stress $T_{xy}$
and second normal stress difference $N_2=T_{yy}-T_{zz}$. In this way,
the key physics will prove robust to choice of model.

We note, however, that both models studied here have a single viscoelastic relaxation time. This is an approximation that should be relaxed in future studies, by adopting multi-mode models.

\subsubsection{Johnson-Segalman model}
\label{sec:JS}

The first model that we consider is the Johnson-Segalman
model~\cite{johnson1977model}, in which the viscoelastic stress
evolves according to
\begin{widetext}
\be
\partial_t\tens{\Sigma}+\vecv{v}.\nabla\tens{\Sigma} = \left(\tsr{\Sigma}\cdot \tsr{\Omega} - \tsr{\Omega}\cdot \tsr{\Sigma} \right) + a\left(\tsr{D} \cdot \tsr{\Sigma} + \tsr{\Sigma} \cdot \tsr{D}\right) + 2G\tens{D} - \frac{1}{\tau}\tsr{\Sigma}+\frac{\ell^2}{\tau}\nabla^2\tens{\Sigma}.
\label{eqn:veceJS}
\ee
\end{widetext}
Here $\tens{D}=\tfrac{1}{2}(\nabla \vecv{v} + \nabla \vecv{v}^T)$ is the symmetric part of the rate of strain tensor $\nabla \vecv{v}_{\alpha\beta}=\partial_\alpha v_\beta$. Its antisymmetric counterpart $\tens{\Omega}=\tfrac{1}{2}(\nabla \vecv{v} - \nabla \vecv{v}^T)$ is the vorticity tensor. The
parameter $a$ describes a relative slip of the viscoelastic component compared with the deformation of the background solvent. It must lie in the range  $-1 \le a\le 1$. The simpler Oldroyd B model is recovered at $a=1$.

In a stationary homogeneous simple shear flow,  the viscoelastic shear stress and second normal stress difference obey the following
functions of the imposed shear rate $\gdot$:
\begin{eqnarray}
  \Sigma_{xy} &=& \frac{\gdot \tau}{1 + (1 - a^2) (\gdot\tau)^2},\nonumber\\
 N_2\equiv \Sigma_{yy}-\Sigma_{zz} &=& \frac{\left(-1 + a\right) \left(\gdot \tau\right)^2}{1 + (1 - a^2) (\gdot\tau)^2}.
  \label{eq:JS_fc}
\end{eqnarray}
For values of the parameters $|a|<1$ and $\etas < 1/8$, the total shear stress $T_{xy}=\Sigma_{xy}+\etas\gdot$ is then a non-monotonic function of the imposed shear rate.  In this regime,
coexisting bands of differing shear rates can form at a common value of the
total shear stress: a phenomenon known as shear banding.  In this work, we consider
only flows that are not shear banded, and so confine ourselves to values of the solvent viscosity $\etas>1/8$. The second normal stress is negative, scaling as $-\gdot^2$ at low shear rates, before saturating to a negative constant at high shear rates. 

The spatial gradient terms prefactored by $\ell$ in Eqn.~\ref{eqn:veceJS} are included in our simulations but do not affect the predictions for edge fracture. (They would be important if the flow were shear banded~\cite{lu2000effects}, as considered in Refs.~\cite{skorski2011loss,hemingway2018edge}.) The microscopic lengthscale $\ell$ is small compared with any bulk lengthscales.

\subsubsection{Giesekus model}
\label{sec:Giesekus}

The second constitutive model that we consider is the Giesekus model~\cite{giesekus1982simple}, in which the viscoelastic stress evolves according to:
\begin{widetext}
\be
\partial_t\tens{\Sigma}+\vecv{v}.\nabla\tens{\Sigma} = \left(\tsr{\Sigma}\cdot \tsr{\Omega} - \tsr{\Omega}\cdot \tsr{\Sigma} \right) + \left(\tsr{D} \cdot\tsr{\Sigma} + \tsr{\Sigma}\cdot \tsr{D}\right) + 2G\tens{D} - \frac{1}{\tau}\tsr{\Sigma}-\frac{\alpha}{\tau}\tsr{\Sigma}\cdot\tsr{\Sigma}+\frac{\ell^2}{\tau}\nabla^2\tens{\Sigma}.
\label{eqn:veceG}
\ee
\end{widetext}
Here $\alpha$ is an anisotropy parameter, which captures an increased
rate of stress relaxation in any regime where the polymer chains are more
strongly aligned. The Oldroyd B model is recovered at $\alpha=0$. In a state of stationary homogeneous simple shear flow, the 
viscoelastic shear and second normal stresses obey~\cite{giesekus1982simple}
\bea
\Sigma_{xy}&=&\frac{(1+N_2)^2\gdot}{1-(1-2\alpha)N_2},\nonumber\\
N_2&=&\frac{\Lambda - 1}{1+(1-2\alpha)\Lambda},
\label{eqn:Giesekus_fc}
\eea
in which
\be
\Lambda^2=\frac{1}{8\alpha(1-\alpha)\gdot^2}\left[ \sqrt{1+16\alpha(1-\alpha)\gdot^2}-1  \right].
\ee
The shear stress $\Sigma_{xy}$ is then a
non-monotonic function of imposed shear rate $\gdot$ for values of the anisotropy parameter $\alpha > 1/2$~\cite{giesekus1982simple}. We shall therefore mostly restrict ourselves to values of $\alpha < 1/2$, again in order to avoid shear-banding. Where we do consider a value of $\alpha>1/2$, we take a large enough solvent viscosity such that the total shear stress $T_{xy}=\Sigma_{xy}+\eta\gdot$ is monotonic, again avoiding banding.
The second normal stress is negative, as in the Johnson-Segalman model: scaling as $-\gdot^2$ at low shear rates, before saturating to a negative constant at high shear rates. 

It is worth remarking that both the Johnson-Segalman and Giesekus models are highly phenomenological. Indeed, although both are widely used in the rheological literature, each has notable pathologies. For example, the Johnson-Segalman model predicts unphysically large and sustained oscillations in the stress in shear startup at high shear rates. However, the spirit of this paper is to derive an instability criterion for edge fracture that is independent of the particular choice of constitutive model, at least in the limit of low strain rates, where both models reduce to a second order fluid as just described. We defer to future work a simulation of more microscopically faithful constitutive models in this geometry: noting, however, that that one such candidate - the Rolie-poly model~\cite{likhtman2003simple} - is excluded by in fact predicting zero second normal stress difference. It would be interesting in future work also to consider the recent tube-based modelling approach of Ref.~\cite{costanzo2018measuring}.

\begin{table*}[!t]
  \def\arraystretch{1.25}
  \resizebox{\linewidth}{!}{%
    \noindent\begin{tabular}{| l | l | l | l | l |}
      \hline
      \textbf{parameter} & \textbf{description} &  \textbf{dimension} & \textbf{value} & \textbf{notes}\\
      \hline
        $L_y$ & channel width & $[L]$ & 1.0 & unit of length\\
            $G$ & polymer modulus & $[G]$ & 1.0 & unit of stress  \\
      $\tau$ & polymer relaxation time & $[T]$ & 1.0 & unit of time  \\
      \hline
        $\gdot$ & applied shear-rate & $[T]^{-1}$ & $10^{-1}\to 10^2$ & important quantity to be varied\\
        $\theta$ & equilibrium contact angle & $[1]$ & $60\degree \to 120\degree$ & important quantity to be varied \\
        $\Gamma$ & surface tension & $[G][L]$ & $0.0 \to 1.0$ & important quantity to be varied\\
          $a$ & slip parameter (JS) & $[1]$ & $0.0\to 0.6$ (usually 0.3) & sets dependencies $\Sigma_{xy}(\gdot), N_2(\gdot)$  \\
      $\alpha$ & anisotropy parameter (Gk) & $[1]$ & $0.1\to 1.0$ (usually 0.4) & sets dependencies $\Sigma_{xy}(\gdot), N_2(\gdot)$ \\
        \hline
         $L_z$ & channel length & $[L]$ & $10.0$ & large aspect ratio $L_z/L_y$\\
      $\Lambda$ & midpoint sample length& $[L]$ & $7.0$ & large enough air gap $(L_z-\Lambda)/L_y$\\
      $\etaa$ & air viscosity  & $[G][T]$ & $0.01$ & small air viscosity $\etaa/G\tau$\\
      $\etas$ & solvent viscosity (JS) & $[G][T]$ & 0.15  & small viscosity ratio $\etas/G\tau$\\
      $\etas$ & solvent viscosity (Gk) & $[G][T]$ & 0.01 ($\alpha\le0.5$); 0.15 ($\alpha>0.5$) & small viscosity ratio $\etas/G\tau$\\
       $\ell$ & polymer microscopic length & $[L]$ & 0.01 & small microscopic length $l/L_y$\\
      $\lm$ & air-polymer interface width & $[L]$ & 0.01 (unless stated otherwise) & small microscopic length $\lm/L_y$\\
      $M$ & molecular mobility & $[L]^2[G]^{-1}[T]^{-1}$ & 0.0001 & rapid phase equilibration\\
      \hline
      $N_y$ & numerical mesh & $[1]$ & large & converge until no dependence\\
       $N_z$ & numerical mesh & $[1]$ & large & converge until no dependence\\
        $Dt$ & numerical timestep & $[1]$ & small & converge until no dependence\\
        \hline
    \end{tabular}
  }
  \caption{Parameters and their dimensions in terms of modulus $[G]$, length $[L]$ and time $[T]$; values used in our numerical simulations; and notes concerning each parameter. The first three parameters listed specify our choice of units. The second five are the key physical parameters to be varied in our study (four within each constitutive model); the set from $L_z$ to $M$ do not affect the key physics, provided each is set to an appropriately large or small value; the final set are numerical parameters and are converged to their appropriate large or small limit. Abbreviations: Johnson-Segalman (JS) and Giesekus (Gk).}
  \label{tbl:params}
\end{table*}

\subsection{Fluid-air coexistence}

In our numerical simulations, we model the coexistence of fluid and air using a phase field approach
with an order parameter $\phi(\vecv{r},t)$, which obeys Cahn-Hilliard
dynamics~\cite{bray1994advances}
\begin{align}
  \partial_t \phi + \vecv{v}.\nabla \phi &= M \nabla^2 \mu.
  \label{eq:SI_CH}
\end{align}
Here $M$ is the molecular mobility, which we assume to be constant. The
chemical potential
\begin{align}
  \mu &= G_\mu \left(-\phi + \phi^3 - \lm^2 \nabla^2 \phi\right),
  \label{eqn:freeEnergy}
\end{align}
in which $G_\mu$ sets the scale of the free energy of
demixing per unit volume.  The functional dependence of this free energy on $\phi$ captures the coexistence of a fluid phase, in which the order parameter $\phi = 1$, with an air phase, in which $\phi = -1$. The elastic modulus $G$ and relaxation time $\tau$ that appear in
the constitutive equations~\ref{eqn:veceJS} and~\ref{eqn:veceG} are then made functions of
$\phi$, with $G(\phi=1)=1$, $\tau(\phi=1)=1$ and $G(\phi=-1)=0$, with $\tau(\phi=-1)=0.002$. In this way, viscoelastic stresses arise only in the fluid phase.

The fluid and air bulk phases are separated by a slightly diffuse interface of thickness
$\lm$,  set by the prefactor to the gradient term in the free energy of Eqn.~\ref{eqn:freeEnergy}. This interfacial thickness is small compared with any bulk lengthscales. Gradients in $\mu$ contribute an additional source term of the form $-\phi \nabla
\mu$ to the force balance condition. This is important in the vicinity of the fluid-air interface, where it confers forces arising from the surface tension of the interface. The surface tension that emerges out of these Cahn-Hilliard dynamics is given by: 
\begin{equation}
  \Gamma = \frac{2 \sqrt{2}}{3} G_\mu \lm.
  \label{eq:SI_surface_tension}
\end{equation}
In having a slightly diffuse fluid-air interface, our
simulations properly capture any motion of the contact line that arises where the fluid-air interface meets the hard walls of the flow cell~\cite{kusumaatmaja2016moving}. We thereby avoid the contact line singularity that would arise if the interface were perfectly sharp.

In our analytical calculations, which are performed in the biperiodic geometry without walls, we shall instead assume the interface between the fluid and air to be infinitely sharp, although still with an equivalent surface tension $\Gamma$. (This sharpness is unimportant in the absence of hard walls, and therefore of any contact line, at least in the early stages of any instability, before bubbles form.) We shall further assume in our analytical calculations that the viscosity of the outside air is negligible compared to that of the fluid. Below we shall demonstrate full agreement between our numerical simulations and analytical calculations, in the biperiodic flow cell, in the physically relevant limit in which the interface is indeed thin compared with any bulk scales, and in which the air viscosity is small compared with that of the fluid.

\subsection{Boundary conditions}
\label{bc}

We adopt periodic boundary conditions in the vorticity direction $z$. Where the fluid meets the hard walls of the flow cell in the $y$ direction, we assume boundary conditions  of no slip and no permeation for the fluid velocity $\vecv{v}$, and zero-gradient  for the viscoelastic stress $\tens{\Sigma}(\vecv{r},t)$:
\be
\vecv{n}\cdot\nabla \,\tens{\Sigma}=0.
\ee
Here $\vtr{n}$ is the outward unit vector normal to the wall. 

For the phase field that captures the fluid-air coexistence, as just described, the boundary
conditions are~\cite{yue2010sharp,Dong2012}
\begin{align}
  \vtr{n}\cdot \nabla \mu &= 0,\\
  \vtr{n}\cdot \nabla \phi &= \frac{-1}{\sqrt{2} \lm} \cos{\theta} \left(1 - \phi^2\right).
  \label{eq:SI_CH_BCs}
\end{align}
In unsheared equilibrium, the contact angle at which the fluid-air
interface meets the hard walls of the flow cell is given by $\theta$. A
value $\theta=90^\circ$ gives a vertical equilibrium interface. A value
$\theta>90^\circ$  gives an interface convex into the air. A value
$\theta<90^\circ$ gives a concave interface. 

The simplified
biperiodic Lees-Edwards geometry has no walls and its equilibrium interface is
always vertical, mimicking the case $\theta=90^\circ$ with walls. The boundary conditions for this case were discussed in Sec.~\ref{sec:geometry} above. 

\subsection{Initial conditions}
\label{sec:ic}

As the initial condition for our simulations/calculations in shear, we take a state of air-fluid coexistence
that has first been equilibrated in the absence of shear. At the end of the equilibration dynamics, just before shear is applied, a small perturbation is
added to the interface's position $h(y)$ along the $z$ direction, $h(y)\to
h(y)+10^{-8}\cos(n\pi y/L_y)$, to trigger edge fracture. We take $n=1$
with walls and $n=2$ in the biperiodic geometry. The viscoelastic stress is initialised so as to lie on the stationary homogeneous constitutive curve, for the shear rate in question. 

\subsection{Nonlinear simulation method}
\label{sec:code}

Each numerical timestep comprises two separate substeps. In the first substep, we enforce the force balance condition in order to update the fluid velocity field $\vtr{v}$ at fixed phase field $\phi$ and polymer stress $\tsr{\Sigma}$. We do so using a streamfunction formulation to ensure incompressible flow.  In the second substep, we update the phase field and viscoelastic stress, with the velocity field fixed. 

The advective terms are implemented using a third order upwinding scheme~\cite{pozrikidis2011introduction}. The spatially local terms in the viscoelastic constitutive equation are updated using an explicit Euler scheme~\cite{press1992numerical}. To implement the spatially diffusive terms, we use a 2D Fourier spectral method in the Lees-Edward biperiodic geometry. With walls present, we use  Fourier modes in the periodic direction $z$, and finite differencing~\cite{press1992numerical} in the flow gradient direction $y$, with wetting conditions implemented using the method described in Ref. \cite{Dong2012}.

As noted above, our simulations explicitly model the coexistence of air and polymeric fluid, using a phase field approach: the phase field $\phi=1$ inside the polymeric fluid and $\phi=-1$ in the outside air.  In addition to the viscoelastic component in the polymeric phase, we also have a background Newtonian viscosity equal to that of the solvent $\eta_s$ inside the polymeric fluid, and a lower value $\eta_a$ in the air. 

To achieve this spatially dependent Newtonian viscosity, we simulate everywhere a Newtonian fluid with the air viscosity $\eta_a$, and further introduce an additional (and nominally viscoelastic) stress tensor ${\boldsymbol{\sigma}}_s$ which obeys Oldroyd-B dynamics with a small enough relaxation time $\tau_s \ll 1$ so as to be essentially Newtonian, and with a (spatially dependent) elastic modulus $G_s(\phi = -1) = 0, G_s(\phi = 1) = (\eta_s - \eta_a) / \tau_s$. This approximates a Newtonian solvent with viscosity $\eta_s = \eta_a + G_s(\phi = 1)\tau_s$ in the polymer phase only and a fluid with viscosity $\eta_a$ in the air phase.

\subsection{Units and parameter values}
\label{sec:units}

\begin{figure}[!b]
  \centering
  \includegraphics[width=0.9\columnwidth]{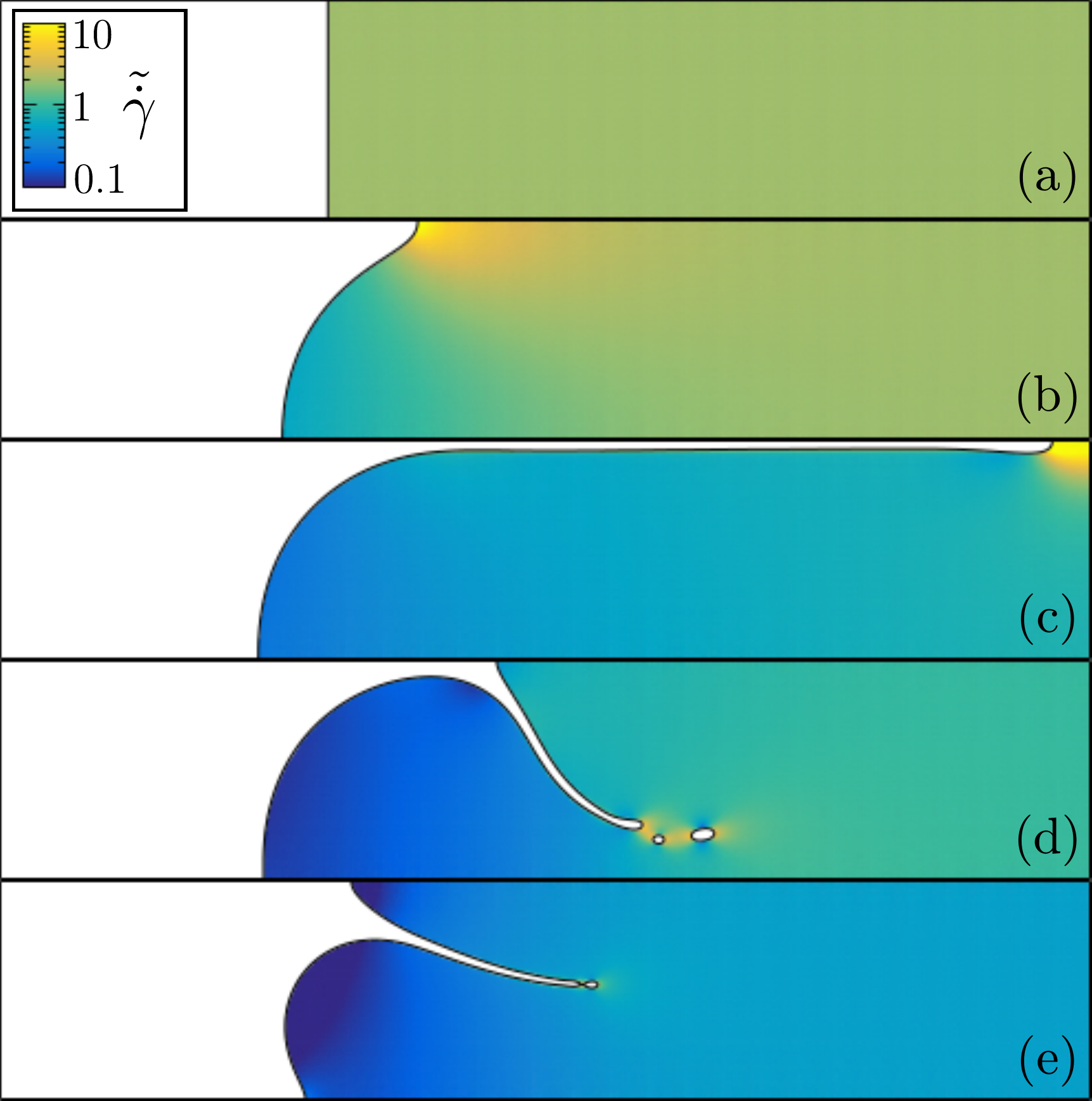}
  \caption{Late-time snapshots corresponding to the state points marked (a) to (e) in Fig.~\ref{fig:PD} (right) for the Giesekus model simulated between hard walls. Color denotes the frame invariant shear rate $\tilde{\gdot} = \sqrt{2\tsr{D}:\tsr{D}}$. Anisotropy parameter $\alpha=0.4$. Equilibrium contact angle $\theta = 90\degree$.}
  \label{fig:giesekus_snapshots}
\end{figure}

\begin{figure*}[!t]
  \includegraphics[width=\textwidth]{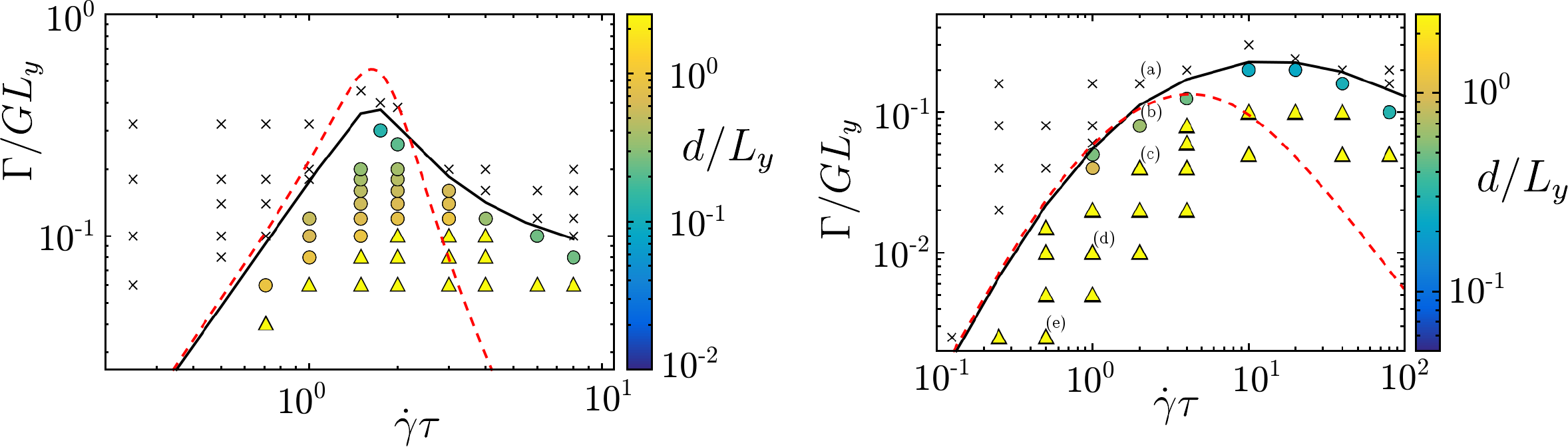}
  \caption{Phase diagram in the plane of dimensionless surface tension and shear rate ($\Gamma / G L_y, \gdot\tau$), for flow in a channel bounded by hard walls.  Crosses show stable states, circles show partially fractured states, and triangles show fully fractured states. Symbol fill colour denotes the interfacial deformation $d$ as defined in the main text.  Black curve shows the numerically measured stability threshold, red dashed curve shows the prediction of this threshold by the linear stability analysis of Sec.~\ref{sec:linear}, which is valid at low strain rates. {\bf Left:} Johnson-Segalman model, $a=0.3$. {\bf Right:} Giesekus model, $\alpha=0.4$. Labels (a) to (e) correspond to snapshots shown in Fig.~\ref{fig:giesekus_snapshots}. Equilibrium contact angle $\theta = 90\degree$ in both cases.}
  \label{fig:PD}
\end{figure*}

The parameters contained in the model equations, geometry and boundary conditions just described are summarised in Table.~\ref{tbl:params}, along with values to be used in our numerics. As can be seen, the four important quantities to be varied in our study are the dimensionless shear rate, $\gdot\tau$, the dimensionless surface tension of the air-fluid interface, $\Gamma/GL_y$, the (obviously dimensionless) equilibrium contact angle $\theta$, and the dimensionless nonlinear parameter, $a$ or $\alpha$, that set the way in which the shear and normal stresses depend on shear rate in either constitutive model. The other parameters do not affect the key physics, provided each takes a large or small value appropriate to the physical limit of interest. For example, we set microscopic lengths small compared to macroscopic lengths. We further set the viscosities of the solvent and air small compared to that of the viscoelastic component. 
Even though the separation of lengthscales (or viscosities) that can feasibly achieved in our simulations is less than that in physical reality, the results are unaffected by this limitation, to within negligible corrections. Again for numerical convenience, the value of air viscosity that we use is much larger than that of actual air, but we have checked that the results obtained do not change upon halving the value used.

\section{Nonlinear simulation results}
\label{sec:nonlinear}

We now present the results of our nonlinear simulations. The basic phenomenon of edge fracture is
exemplified by the five snapshots shown in
Fig.~\ref{fig:giesekus_snapshots} at some long time $t$ after the switch-on of shear at time $t=0$, for the Giesekus model simulated between hard walls.  At any given imposed strain rate, an
air-fluid interface with a high surface tension is undisturbed by the
flow and retains its equilibrium shape, as in snapshot (a). (We shall denote
such undisturbed states by black crosses in the phase diagram Fig.~\ref{fig:PD}, which we discuss below.) For intermediate values of  the interfacial
surface tension, the interface partially fractures, displacing in the
$z$ direction by a distance $O(L_y)$ that is set by the gap between the rheometer
plates in the $y$ direction, before finally settling to a new steady state
shape, different from its unsheared equilibrium one. See snapshot (b) in Fig.~\ref{fig:giesekus_snapshots}. (We denote such partially fractured
states by circles in Fig.~\ref{fig:PD}.) For low values of the interfacial surface tension, the
interface fully fractures, displacing in the $z$ direction a distance
$O(\Lambda)$ set by the sample width in that direction. See snapshots (c)-(e) in Fig.~\ref{fig:giesekus_snapshots}. (We denote such fully fractured states by
yellow closed triangles in Fig.~\ref{fig:PD}.) In such cases, the
system never attains a new steady state. Instead, the sample may (for example) completely de-wet either
wall, and/or air bubbles may invade the fluid. Which of these happens in practice depends on the wetting
angle in a way that we shall investigate further below.

In Fig.~\ref{fig:PD}, we collect into a phase diagram the results of
many such simulations across a full range of values of the surface
tension and imposed shear rate, for both the Johnson-Segalman model
(left panel) and the Giesekus model (right panel).  The symbol shapes
(crosses, circles, triangles) are as described in the previous
paragraph. The state-points corresponding to the snapshots of
Fig.~\ref{fig:giesekus_snapshots} for the Giesekus model are indicated
by letters (a)-(e) in the right panel. As can be seen, broadly the
same phase behaviour arises in both models, with stability at high
values of interfacial surface tension, and instability for low surface
tension. At any fixed (low) value of surface tension, we find
stability for weak flows $\gdot\to 0$, as expected, followed by a
window of instability for intermediate values of the shear rate, and
finally a regime of re-entrant stability at high strain rates. We
shall discuss this re-entrance in more detail below. It arises from a
saturation (in both constitutive models) of the growth of $|N_2|$ with
shear rate $\gdot$, which may not be physically realistic. Even if the
re-entrant regime does in principle exist in real fluids, it may be
hard to access in experimental practice: one would either need to
sweep the shear rate upward through the regime of shear rates where
edge fracture does arise, or alternatively perform a shear startup at
a high shear rate, which is in itself likely to be unstable. From
an experimental viewpoint, it is likely that polymers melts mostly lie
in the unstable regime of Fig.~\ref{fig:PD}, at all but the lowest
strain rates; whereas some polymer solutions, with their lower modulus
$G$, may be in the stable regime.

\begin{figure*}
  \centering
  \includegraphics[width=\textwidth]{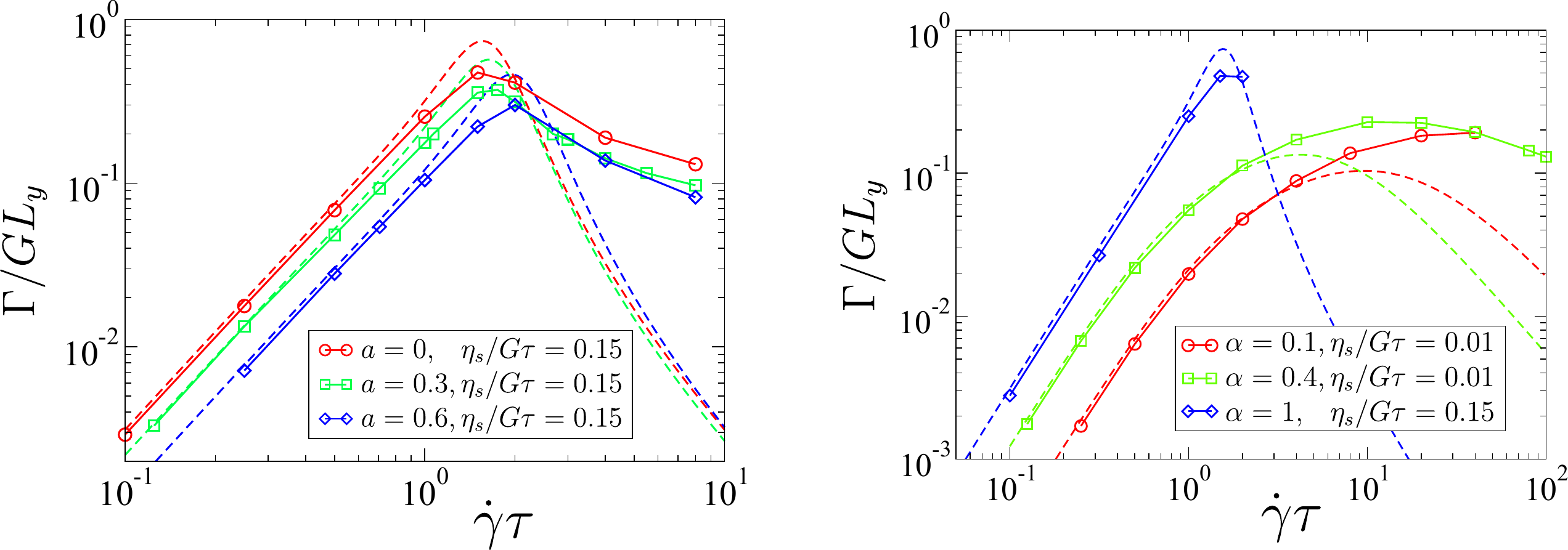}
  \caption{Thresholds for onset of the edge fracture instability for a fluid sheared between hard wells, for an equilibrium contact angle $\theta=90\degree$. Solid lines: thresholds measured from the early time dynamics of the full nonlinear simulations. Dashed lines: thresholds calculated from the linear stability analysis in Sec.~\ref{sec:linear}, valid for small shear rates. {\bf Left:} Johnson-Segalman model, for several values of the slip parameter $a$. {\bf Right:} Giesekus model, for several values of the anisotropy parameter $\alpha$.}
  \label{fig:JS_vary_a}
\end{figure*}

The colorscale of the symbol fill in Fig.~\ref{fig:PD}, in parameter regimes where (either partial or full) edge fracture arises, denotes the degree to which the interface has deformed at long times after the switch-on of flow at time $t=0$. This is defined as follows. We denote by $h(y,t)$ the location along $z$ of the interface at height $y$ across the gap at time $t$. In the sheared state at any time $t$, ignoring any bubbles that form, we subtract the leftmost position of the interface from the rightmost position to give $D(t)={\rm max}(h(y,t))-{\rm min}(h(y,t))$.  From this we subtract the value of $D(t=0)$ in the initial unsheared state. (For an equilibrium contact angle $\theta=90\degree$, $D(0)=0$.) The difference between these, normalised by the gap width $L_y$, gives the degree of interfacial deformation in shear, $d(t)=[D(t)-D(0)]/L_y$.  The values of this quantity shown in yellow in the triangles of Fig.~\ref{fig:PD} should be interpreted as lower bounds, because no steady state is reached as $t\to\infty$ in such cases.

The black solid line in Fig.~\ref{fig:PD} shows the threshold between a fully stable interface (states shown by crosses) and an interface that partially fractures (states shown by circles). This is calculated by measuring, in the simulation performed at each  value of $\Gamma,\gdot$, the weight in the Fourier modes $\exp(iqy)$ of the interfacial displacement $h(y,t)-h(0,t)$ as a function of time $t$. In the unstable regime (below the black line), this (initially) grows exponentially for some modes; in the stable regime (above the black line), it decays exponentially for all modes, from our initially small perturbation $O(10^{-8})$. Plotting the growth (or decay) constant (ie, the eigenvalue) of the most dangerous mode as a function of surface tension $\Gamma$ at any fixed $\gdot$ enables us to find the value of $\Gamma(\gdot)$ at which the interface is neutrally stability, with an eigenvalue of zero for the most dangerous mode. This gives the threshold shown by the black solid line. The red dashed line is the analytical prediction given by the linear stability calculation of Sec.~\ref{sec:linear} below, which is performed in the limit of low strain rate.

We emphasise that Fig.~\ref{fig:PD} makes quantitative predictions for the onset of edge fracture as a function of shear rate and surface tension that are testable experimentally, and we hope that this will motivate future experimental work along these lines.

So far, we have explored the phase behaviour across a wide range of values of surface tension and imposed shear rate, within the Johnson-Segalman model  for one fixed value of the slip parameter $a$, and within the Giesekus model for one fixed value of the anisotropy parameter $\alpha$.  We now explore the dependence of this phase behaviour on the value of $a$ (in Johnson-Segalman) and $\alpha$ (in Giesekus). This is shown in Fig.~\ref{fig:JS_vary_a}. In each case, the neutral stability curve extracted from the nonlinear simulations in the way described in the previous paragraph is shown as a solid curve, and the prediction of the linear stability analysis of Sec.~\ref{sec:linear} is shown by the dashed curve. 
The results for the Johnson-Segalman model are given in the left panel. As can be seen, the unstable regime becomes more limited for large values of $a$. This is consistent with a scenario in which the edge fracture instability is driven by the second normal stress $N_2$: the Johnson-Segalman model recovers the Oldroyd B model in the limit $a\to 1$, with zero second normal stress $N_2(\gdot,a=1)=0$ (recall   Eqn.~\ref{eq:JS_fc}), but non-zero first normal stress.   The results for the Giesekus model are shown in the right panel. The unstable regime again becomes more limited for small values of $\alpha$, consistent with the Oldroyd B model being recovered for $\alpha= 0$. 

\begin{figure}[!b]
  \centering
  \includegraphics[width=\columnwidth]{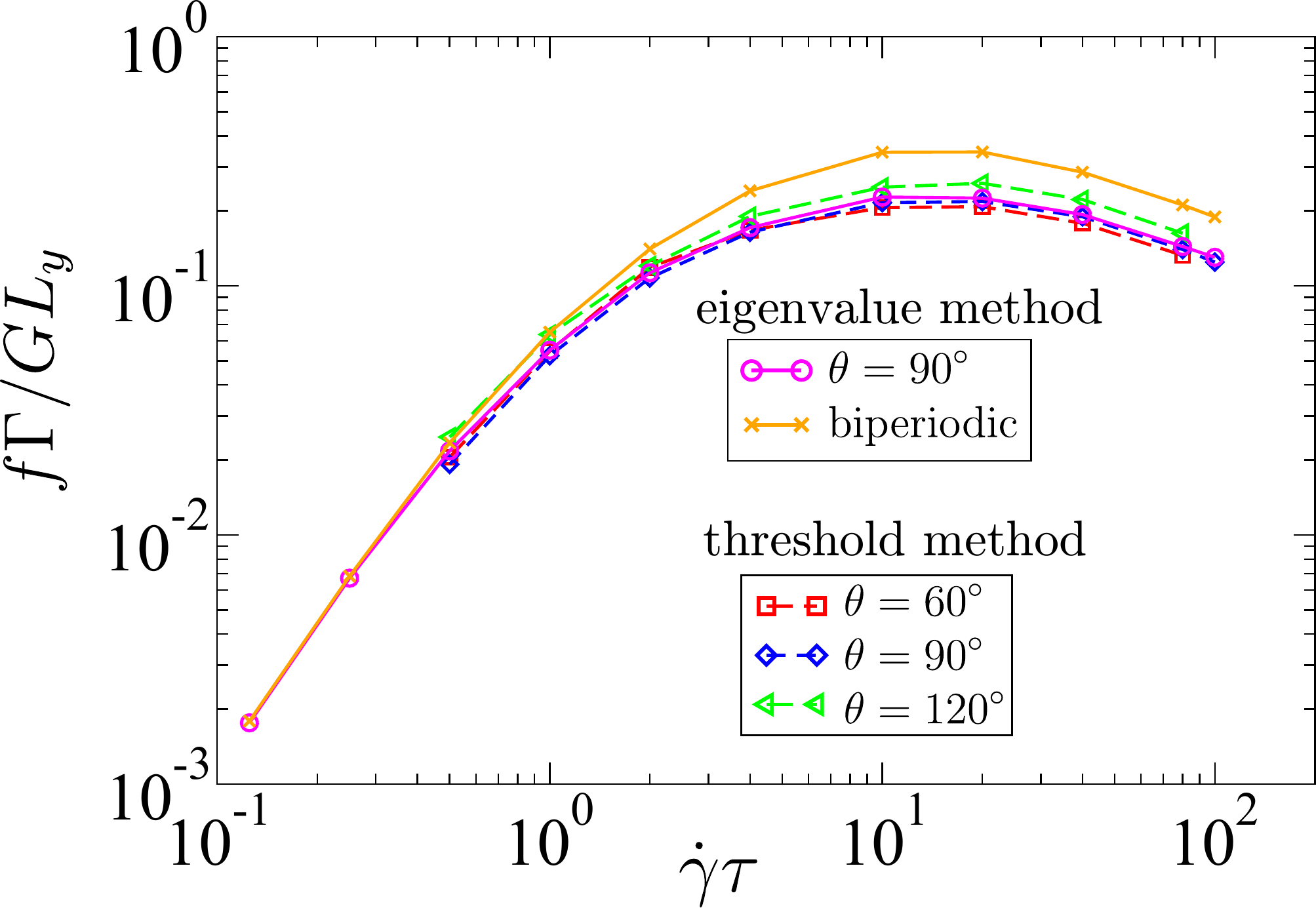}
  \caption{Threshold for the onset of edge fracture within the Giesekus model for several different wetting conditions (for simulations performed in a channel with hard walls) and for the two different kinds of boundary condition (hard walls versus a biperiodic flow cell). Note that the minimum possible wavenumber $q_{\rm min}$ depends on the boundary condition: $q_{\rm min} = \pi / L_y$ for hard walls and $q_{\rm min} = 2 \pi / L_y$ in the biperiodic geometry.  Accordingly, the ordinate in this figure is scaled by a factor $f=2$ in the biperiodic simulations. ($f=1$ for the simulations between hard walls.)  Anisotropy parameter $\alpha = 0.4$.}   
  \label{fig:robust}
\end{figure}

The simulation results shown so far have all been for a value of the equilibrium contact angle $\theta=90\degree$, for which the interface is initially flat.
To check for robustness with respect to this choice of boundary
condition, we now compare the threshold for the onset
of edge fracture for different values of the
equilibrium contact angle, for a channel with hard walls. Recall that for the particular case of $\theta=90\degree$, we calculated the threshold by identifying the value of surface tension (at any fixed imposed strain rate) at which the eigenvalue of the most dangerous mode crosses zero. A method of determining the threshold that applies more easily across values of $\theta\neq 90\degree$ is instead to identify the value of surface tension (at any fixed imposed shear rate) below which the degree of interfacial deformation, $d$, first exceeds $0.1$. (Good agreement is obtained between these two methods at $\theta=90\degree$; the eigenvalue method is only used at $\theta=90\degree$.) As seen in Fig.~\ref{fig:robust}, the threshold for the onset of edge fracture is indeed robust (to within small corrections) against variations of the equilibrium contact angle. 

\begin{figure}[!t]
  \includegraphics[width=0.45\textwidth]{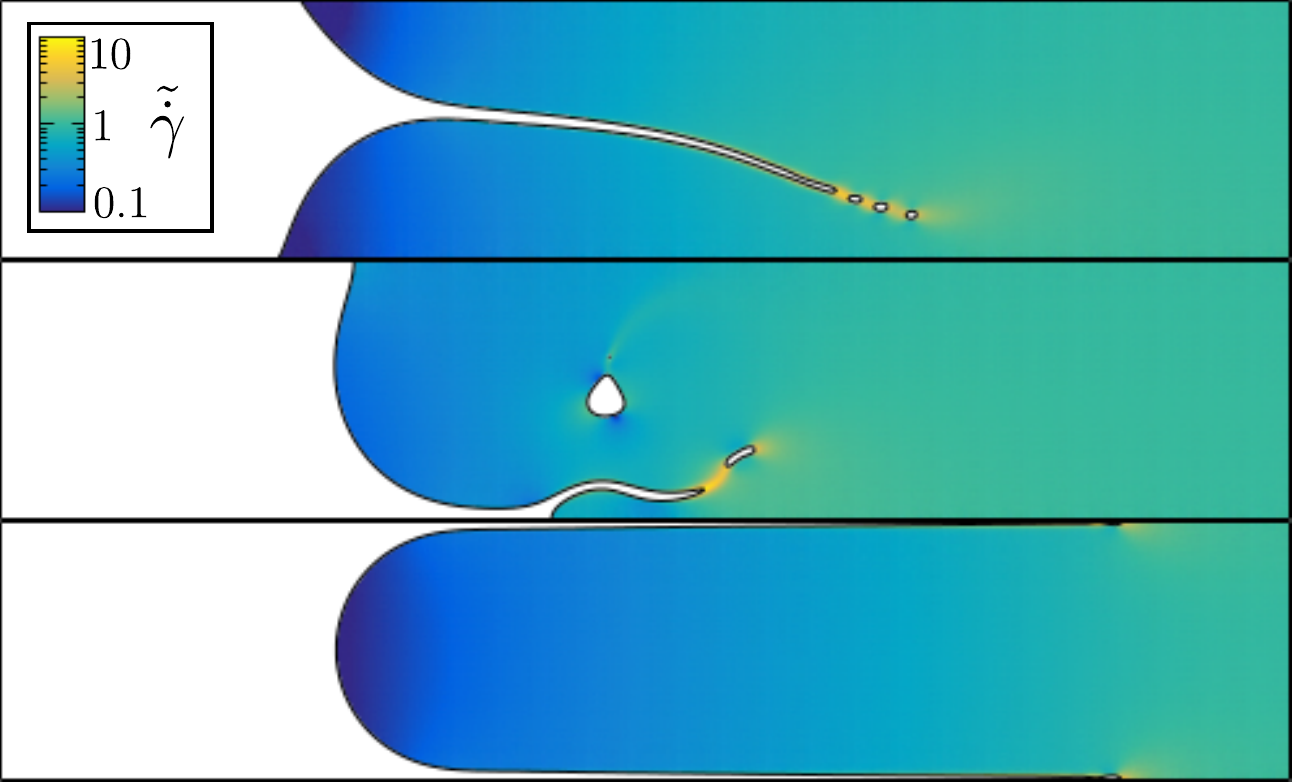}
  \caption{Effect of the equilibrium contact angle on the mode of the edge fracture instability, within the Giesekus model. Color denotes the frame invariant shear rate $\tilde{\gdot} = \sqrt{2\tsr{D}:\tsr{D}}$. Anisotropy parameter $\alpha=0.4$. Surface tension $\Gamma / G L_y = 0.005$. Imposed shear rate $\gdotb \tau =1.0$. Contact angle $\theta = 60\degree, 90\degree, 120\degree$  in the top, middle, and bottom  snapshots.}
  \label{fig:nonlinear_TODO}
\end{figure}

Despite this robustness in the onset threshold, the spatial mode of instability does vary with the equilibrium contact angle, particularly for values of the surface tension deep inside the unstable regime. This can be seen in the snapshots of Fig.~\ref{fig:nonlinear_TODO}. For values of the contact angle $\theta<90\degree$, for which the fluid tends to wet the walls, the air first invades the fluid at a central
location across the channel width $y$, leading to the formation of bubbles within the sample's bulk. For values $\theta>90\degree$, the air instead invades along the channel walls, often leading the fluid eventually to de-wet the walls entirely. (The air from the left side of the cell shown in Fig.~\ref{fig:nonlinear_TODO} finally joins up with that from the right side, not shown.) 

We note that larger angles, such as $\theta=120\degree$, correspond to an experimental case where a polymer sample is squeezed between a cone and plate without trimming. In contrast, $\theta=90\degree$ corresponds to the ideal case where the surface of the sample is part of a sphere. Therefore, results for experimental papers on melts are likely to fall in between these two cases. Indeed, the results in Fig.~\ref{fig:nonlinear_TODO} corroborate the experimental observations of Fig. 3 of~\cite{schweizer2008departure}, in which a higher shear layer was seen near the walls for an initially convex sample profile, whereas a higher shear layer was seen mid-gap for an initially concave sample profile.

All the simulation results presented so far have been for a fluid sheared between hard walls. We finally compare those results to the corresponding results obtained in a biperiodic flow cell. As seen in Fig.~\ref{fig:robust},  good agreement is obtained between these two cases, particularly at low strain rates $\gdot\to 0$. Note that this agreement is obtained after rescaling the box size $L_y$ by a factor $2$ (as discussed further in the caption of Fig.~\ref{fig:robust}) to allow for the (relatively trivial) fact that the minimum possible wavevector $q_{\rm min}=\pi/L_y$ with hard walls, whereas $q_{\rm min}=2\pi/L_y$ in the biperiodic geometry. This will enable us to perform a linear stability analysis for the onset of edge fracture in the simpler geometry of the biperiodic flow cell, for which the eigenfunctions are harmonic functions of $y$. It is to this linear analysis that we now turn.

\section{Linear stability analysis}
\label{sec:linear}

We now perform a linear stability
analysis to derive a criterion for the onset of edge fracture. This calculation is performed
in the limit of low strain rates within the biperiodic flow geometry. We now take the viscosity of the air outside the fluid to be zero (it was small but non-zero in our nonlinear simulations), and assume the fluid-air interface to be infinitely thin (it had a slightly diffuse width in our nonlinear simulations), although still with a finite surface tension $\Gamma$. Readers who do not wish to follow the details can skip directly to the result in Eqn.~\ref{eqn:criterion}.

\subsection{Governing equations}

We first recall for convenience the governing equations.

\subsubsection{Force balance in the fluid bulk}

Inside the fluid bulk, the flow obeys the incompressibility condition
\be
\nabla.\vecv{v}=0,
\label{eqn:incomp1}
\ee
together with the force balance condition
\be
\nabla\cdot\tens{T}=0,
\label{eqn:fb1}
\ee
in which the total stress
\be
\tens{T}=2\etas\tens{D}+\tens{\Sigma}-p\tens{I}.
\label{eqn:stress1}
\ee

\subsubsection{Bulk viscoelastic constitutive equation}

The viscoelastic stress obeys a constitutive equation
\begin{widetext}
\be
\partial_t\tens{\Sigma}+\vecv{v}.\nabla\tens{\Sigma} = \left(\tsr{\Sigma} \cdot\tsr{\Omega} - \tsr{\Omega}\cdot \tsr{\Sigma} \right) + a\left(\tsr{D}\cdot \tsr{\Sigma} + \tsr{\Sigma}\cdot \tsr{D}\right) + 2G\tens{D} - \frac{1}{\tau}\tsr{\Sigma}-\frac{\alpha}{\tau G}\tsr{\Sigma}\cdot\tsr{\Sigma},
\label{eqn:vece1}
\ee
\end{widetext}
with $-1<a<1$ and $\alpha=0$ in the Johnson-Segalman model, and $a=1$ and $0<\alpha\le 1$ in the Giesekus model. Making the twin assumptions that both the time derivative and the advective derivative on the left hand side of Eqn.~\ref{eqn:vece1} can be neglected (we shall return below to justify these assumptions), we find that for slow flows, $\gdot \tau < 1$, the solution to Eqn.~\ref{eqn:vece1} can be written as an expansion in increasing powers of the flow rate tensors $\tens{D},\tens{\Omega}$:
\begin{widetext}
\be
\tens{\Sigma}=2\tens{D}G\tau+2(\tens{D}\cdot\tens{\Omega}-\tens{\Omega}\cdot\tens{D})G\tau^2 +2(a-\alpha)(\tens{D}\cdot\tens{D}+\tens{D}\cdot\tens{D})G\tau^2+ \rm{h. o. t.}
\label{eqn:second_order}
\ee
\end{widetext}
We ignore the higher order terms (h.o.t.) in what follows. In this
case, the Johnson-Segalman and Giesekus models both reduce to this
so-called `second-order fluid'. Note that it was identified in
Ref.~\cite{porteous1972linear} that the second order fluid contains an
unphysical mode of instability because of the inherent instability of the rest state and
incompatibility with the second law of thermodynamics if the disturbance timescale is not
large compared to the characteristic relaxation timescale of the
fluid. Because we are using the second order fluid in our analytical
work to calculate the boundary of instability onset, where the
disturbance timescale is by definition infinity, our use of the second
order fluid is deemed safe in this regard. We further note that the
analytical results obtained within this second order fluid calculation
agree quantitatively with our numerical simulations of the
Johnson-Segalman and Giesekus models at low strain rates (and
qualitatively at high strain rates).

\subsubsection{Force balance across fluid-air interface}

The condition of force balance across the interface between the fluid and outside air is given by 
\be
\vecv{n}\cdot\tens{T}+\Gamma\vecv{n}\nabla_{\rm int}\cdot\vecv{n}=0,
\label{eqn:interface1}
\ee
where $\vecv{n}$ is the interface normal into the fluid and $\nabla_{\rm int}$ is the interfacial gradient operator. (In the diffuse interface case, this condition of interfacial force balance emerges naturally from having a term of the form $\phi\nabla\mu$ in the bulk force balance equation. In the sharp interface limit that we consider in this analytical calculation, the interfacial condition provided by Eqn.~\ref{eqn:interface1} must be imposed separately.)

\subsubsection{Motion of fluid interface}

Finally, the position $h(y)$ along the $z$ axis of the interface at any height $y$ across the flow cell moves with the $z-$component of fluid velocity as
\be
\partial_t h=v_z.
\label{eqn:interface1b}
\ee

\subsection{Base state and small perturbation}

To perform a linear stability analysis, we represent the state of the system by an underlying 
homogeneous base state, denoted by subscript $0$. This base state is stationary, in the sense of being a time-independent solution of the model equations, but not (necessarily) stable against the onset of edge fracture. It corresponds to a state of stationary shear in an initially unfractured sample, in which the interface is
flat and the flow uniform.  To this, we add a small perturbation, denoted by over-tildes, which represents the precursor of edge fracture.

Accordingly, in the fluid bulk we write the velocity field
\be
\vecv{v}=\vecv{v}_0+\tilde{\vecv{v}}=\left(\begin{matrix}
    \gdot_0 y  \\
    0 \\
    0  
\end{matrix}\right)
+
\left(\begin{matrix}
   \tilde{v}_x\\
   \partial_z\tilde{\psi}\\
   -\partial_y\tilde{\psi}
   \end{matrix}\right),
   \label{eqn:vbp}
\ee
in which $\tilde{v}_x=\tilde{v}_x(y,z,t)$ and $\tilde{\psi}=\tilde{\psi}(y,z,t)$.
In restricting the velocity components to depend only on $y$ and $z$, and in writing the $y$ and $z$ velocity components as the $y-z$ curl of a streamfunction $\psi$, we automatically ensure that the incompressibility condition of Eqn.~\ref{eqn:incomp1} is obeyed. 

The strain rate tensor follows from Eqn.~\ref{eqn:vbp} as:
\begin{widetext}
\be
\nabla\vecv{v}=\nabla\vecv{v}_0+\nabla\tilde{\vecv{v}}=
\left(\begin{matrix}
    0      & 0 & 0 \\
    \gdot_0       & 0 & 0 \\
    0      & 0 & 0
    \end{matrix}\right)
    +
    \left(\begin{matrix}
    0      & 0 & 0 \\
    \partial_y\tilde{v}_x       & \partial_y\partial_z\tilde{\psi} & -\partial_y^2\tilde{\psi} \\
    \partial_z\tilde{v}_x     & \partial_z^2\tilde{\psi} & -\partial_y\partial_z\tilde{\psi}
    \end{matrix}\right).
    \label{eqn:gradvbp}
\ee
We similarly write the viscoelastic stress tensor as the sum of a homogeneous base state plus small perturbations:
\be
\tens{\Sigma}=\tens{\Sigma}_0+\tilde{\tens{\Sigma}}=
\left(\begin{matrix}
    \Sigma_{xx0}(\gdot_0)      & \Sigma_{xy0}(\gdot_0) & 0 \\
    \Sigma_{xy0}(\gdot_0)       & \Sigma_{yy0}(\gdot_0) & 0 \\
    0      & 0 & \Sigma_{zz0}(\gdot_0)
    \end{matrix}\right)
    +
    \left(\begin{matrix}
    \tilde{\Sigma}_{xx}      & \tilde{\Sigma}_{xy}  & \tilde{\Sigma}_{xz}  \\
    \tilde{\Sigma}_{xy}       & \tilde{\Sigma}_{yy}  & \tilde{\Sigma}_{yz}  \\
    \tilde{\Sigma}_{xz}    & \tilde{\Sigma}_{yz}  & \tilde{\Sigma}_{zz} 
    \end{matrix}\right),
    \label{eqn:stressbp}
\ee
\end{widetext}
in which $\tilde{\Sigma}_{ij}=\tilde{\Sigma}_{ij}(y,z,t)$. The total stress tensor $\tens{T}=\tens{T}_0+\tilde{\tens{T}}$ follows an analogous componentwise format, which we do not write out. 

Our strategy now is to substitute  Eqns.~\ref{eqn:vbp} to~\ref{eqn:stressbp} into the governing  Eqns.~\ref{eqn:incomp1} to~\ref{eqn:interface1}, and expand to first order in the amplitude of the perturbations. This will result in a set of linearised equations that govern the dynamics of the perturbations, the solution of which will allow us to delineate the regime in which the perturbations grow to give edge fracture.

\subsection{Linearised equations of motion}

\subsubsection{Linearised force balance condition in the fluid bulk}

The conditions of force balance and incompressibility, Eqns.~\ref{eqn:incomp1} to~\ref{eqn:stress1}, are linear in the quantities they contain. Therefore, the linearised
 force balance condition is simply $\nabla.\tilde{\tens{T}}=0$. This can be written componentwise as: 
\bea
0&=&\partial_y \tilde{T}_{xy}+\partial_z\tilde{T}_{xz},\nonumber\\
0&=&\partial_y\partial_z(\tilde{T}_{yy}-\tilde{T}_{zz})+(\partial_z^2-\partial_y^2)\tilde{T}_{yz}.
\label{eqn:bulk}
\eea
The first of these is 
the $x$ component of force balance. The second is the curl in the $y-z$ plane of the $y$ and $z$
components of force balance. Writing the perturbed total stress as the sum of the perturbed Newtonian stress, the perturbed viscoelastic stress, and the perturbed pressure, $\tilde{\tens{T}}=2\eta\tilde{\tens{D}}+\tilde{\tens{\Sigma}}-\tilde{p}\,\tens{I}$, we can exactly rewrite  Eqns.~\ref{eqn:bulk} as:
\bea
0&=&\etas\nabla^2\tilde{v}_x+\partial_y \tilde{\Sigma}_{xy}+\partial_z\tilde{\Sigma}_{xz},\nonumber\\
0&=&\etas\nabla^4\tilde{\psi}+\partial_y\partial_z(\tilde{\Sigma}_{yy}-\tilde{\Sigma}_{zz})+(\partial_z^2-\partial_y^2)\tilde{\Sigma}_{yz},
\label{eqn:bulk1}
\eea
in which the 2D Laplacian $\nabla\equiv\partial_y^2+\partial_z^2$.
(The pressure does not appear in Eqn.~\ref{eqn:bulk1} because $\partial_x \tilde{p}=0$ in the first equation and the $yz$ curl of $\nabla \tilde{p}$ is zero in the second.)

Substituting into Eqn.~\ref{eqn:bulk1} the componentwise forms of $\tilde{\Sigma}_{ij}$ that will be set out in the next subsection, we find finally the linearised bulk equations:
\bea
0&=&(G\tau+\etas)\nabla^2v_x+G\tau^2(a-\alpha)\gdot_0\nabla^2\partial_z\psi,\nonumber\\
0&=&(G\tau+\etas)\nabla^4\psi+G\tau^2(a-\alpha-1)\gdot_0\nabla^2\partial_zv_x.
\label{eqn:bulk_linear}
\eea

\subsubsection{Linearised bulk viscoelastic constitutive equation}

For parameter values in the vicinity of the threshold of onset of instability in the phase diagram (as in  Fig.~\ref{fig:PD}, for example), edge fracture develops only slowly. This allows us to neglect the term $\partial_t\tens{\Sigma}$ in the viscoelastic constitutive equation. At the level of terms that are linear in the perturbation, the advective term, $\vecv{v}\cdot\nabla\tens{\Sigma}$, is also negligible, because the perturbed quantities vary only in the $y-z$ plane, which is orthogonal to the velocity of the base flow. For imposed shear rates $\gdot_0\tau\ll 1$, this enables us to work with the form of the viscoelastic constitutive equation given by Eqn.~\ref{eqn:second_order}. Expanding this to linear order in the amplitude of the perturbations, we get
\begin{widetext}
\be
\frac{\tilde{\tens{\Sigma}}}{G}=2\tilde{\tens{D}}\tau+4\tau^2\left[(a-\alpha)\tens{D}_0\cdot\tilde{\tens{D}}+(a-\alpha)\tilde{\tens{D}}\cdot\tens{D}_0+\tens{D}_0\cdot\tilde{\tens{\Omega}}+\tilde{\tens{D}}\cdot\tens{\Omega}_0\right]^{\rm S}.
\label{eqn:linvece}
\ee
\end{widetext}
Here we use the notation that $\tens{A}^{\rm S}=\tfrac{1}{2}(\tens{A}+\tens{A}^{\rm T})$, for any tensor $\tens{A}$. Componentwise, Eqn.~\ref{eqn:linvece} can be written as:
\begin{widetext}
\be
\frac{\tilde{\tens{\Sigma}}}{G}=
  \tau\left(\begin{matrix}
    0      & \partial_y\tilde{v}_x & \partial_z\tilde{v}_x  \\
    \partial_y\tilde{v}_x       & 2\partial_y\partial_z\tilde{\psi} & (\partial_z^2-\partial_y^2)\tilde{\psi}\\
    \partial_z\tilde{v}_x     & (\partial_z^2-\partial_y^2)\tilde{\psi} & -2\partial_y\partial_z\tilde{\psi}
    \end{matrix}\right)
    +(a-\alpha)\gdot_0\tau^2
    \left(\begin{matrix}
    2  \partial_y\tilde{v}_x    & 2\partial_y\partial_z\tilde{\psi}  & (\partial_z^2-\partial_y^2)\tilde{\psi}  \\
    2\partial_y\partial_z\tilde{\psi}       & 2  \partial_y\tilde{v}_x & \partial_z\tilde{v}_x  \\
    (\partial_z^2-\partial_y^2)\tilde{\psi}   & \partial_z\tilde{v}_x  & 0
    \end{matrix}\right)
    +\gdot_0\tau^2
     \left(\begin{matrix}
    2  \partial_y\tilde{v}_x    & \partial_y\partial_z\tilde{\psi}  & -\partial_y^2\tilde{\psi}  \\
    \partial_y\partial_z\tilde{\psi}       & -2  \partial_y\tilde{v}_x & -\partial_z\tilde{v}_x  \\
     -\partial_y^2\tilde{\psi}   & -\partial_z\tilde{v}_x  & 0
    \end{matrix}\right).
\label{eqn:linvece_components}
\ee
\end{widetext}

\subsubsection{Linearised force balance across air-fluid interface}

In similar spirit, we write the position along the $z$ axis of the interface at any
location $y$ across the gap as $h_0+\tilde{h}(y,t)$. We further choose
the origin of the $z$ axis to lie at the location of the (unperturbed) interface, so that $h_0=0$, with the positive $z$ direction defined so as to have
fluid for $z>0$ and air for $z<0$. To first order, the interface normal $\hat{\vecv{n}}$ in Eqn.~\ref{eqn:interface1} is then written $\hat{\vecv{n}}=\hat{\vecv{z}}-\partial_y \tilde{h} \hat{\vecv{y}}$. In linearised form, the interfacial force balance condition, Eqn.~\ref{eqn:interface1}, is then written componentwise as:
%
\bea
0&=&\tilde{T}_{xz}|_{z=0^+}-\Delta\sigma\,\,\partial_y \tilde{h},\nonumber\\
0&=&\tilde{T}_{yz}|_{z=0^+}- N_2 \,\partial_y \tilde{h},\nonumber\\
0&=&\tilde{T}_{zz}|_{z=0^+}+\Gamma\,\partial_y^2\tilde{h}.
\label{eqn:interfaceb}
\eea
Here we have denoted by $N_2$ the second normal stress difference in the base state, $\Sigma_{yy0}-\Sigma_{zz0}$, and by $\Delta\sigma$ the shear stress in the base state, $T_{xy0}=\Sigma_{xy0}+\etas\gdot_0$. We use the notation $\Delta\sigma$ rather than simply $\sigma$, because the relevant quantity is the {\em jump} in shear stress between the fluid and outside air. (In this linear calculation we are however assuming the stress in the outside air to be zero, so $\Delta\sigma=\sigma$. $N_2$ is always zero in the outside air, with no $\Delta$ notation needed in that case.) The notation $\tilde{T}_{ij}|_{z=0^+}$ denotes the $ij$-th component of the perturbation to the polymer stress immediately next to the interface, just inside the polymer phase.

While it may seem counter intuitive to discuss a jump in the shear stress across an interface, it is important to emphasise that by shear stress we mean $\Sigma_{xy}$, consistent with the main flow direction being $\xhat$ and flow gradient direction being $\yhat$. It is this quantity that jumps across the interface, which has its normal in the vorticity direction $\zhat$. Because of the mutually different directions involved, a jump in $\Sigma_{xy}$ across a (perfectly flat) interface with normal along $\zhat$ does not violate the condition of force balance across the interface.

Inserting into Eqn.~\ref{eqn:interfaceb} the componentwise form of the perturbation to the total stress, $\tilde{T}_{ij}=\tilde{\Sigma}_{ij}+2\etas\tilde{D}_{ij}-\tilde{p}\delta_{ij}$, and eliminating the perturbation to the pressure $\tilde{p}$ via use of the linearised bulk force balance equation, we find the final linearised condition of force balance across the fluid-air interface:
\begin{widetext}
\bea
0&=&\left[(G\tau+\etas)\partial_z\tilde{v}_x-G\tau^2(1+a-\alpha)\gdot_0\partial_y^2\tilde{\psi}+aG\tau^2\gdot_0\partial_z^2\tilde{\psi}\right]|_{z=0^+}-\Delta\sigma\,\,\partial_y\tilde{h},\nonumber\\
0&=&\left[(G\tau+\etas)(\partial_z^2-\partial_y^2)\tilde{\psi}+G\tau^2(a-\alpha-1)\gdot_0\partial_z\tilde{v}_x\right]|_{z=0^+}- N_2 \,\partial_y\tilde{h},\nonumber\\
0&=&\left[-(G\tau+\etas)(3\partial_y^2+\partial_z^2)\partial_z\tilde{\psi}+2\gdot_0(1-a-\alpha)G\tau^2\partial_y^2\tilde{v}_x+\gdot_0(1-a-\alpha)G\tau^2\partial_z^2\tilde{v}_x\right]|_{z=0^+}+\Gamma\,\partial_y^3\tilde{h}.
\label{eqn:interfacec}
\eea
\end{widetext}

\subsubsection{Linearised equation of interfacial motion}

The linearised form of the equation of interface motion, Eqn.~\ref{eqn:interface1}, is:
\be
\partial_t\tilde{h}=-\partial_y\tilde{\psi}|_{z=0^+}.
\label{eqn:moves}
\ee

\begin{figure}[!b]
  \centering
  \includegraphics[width=\columnwidth]{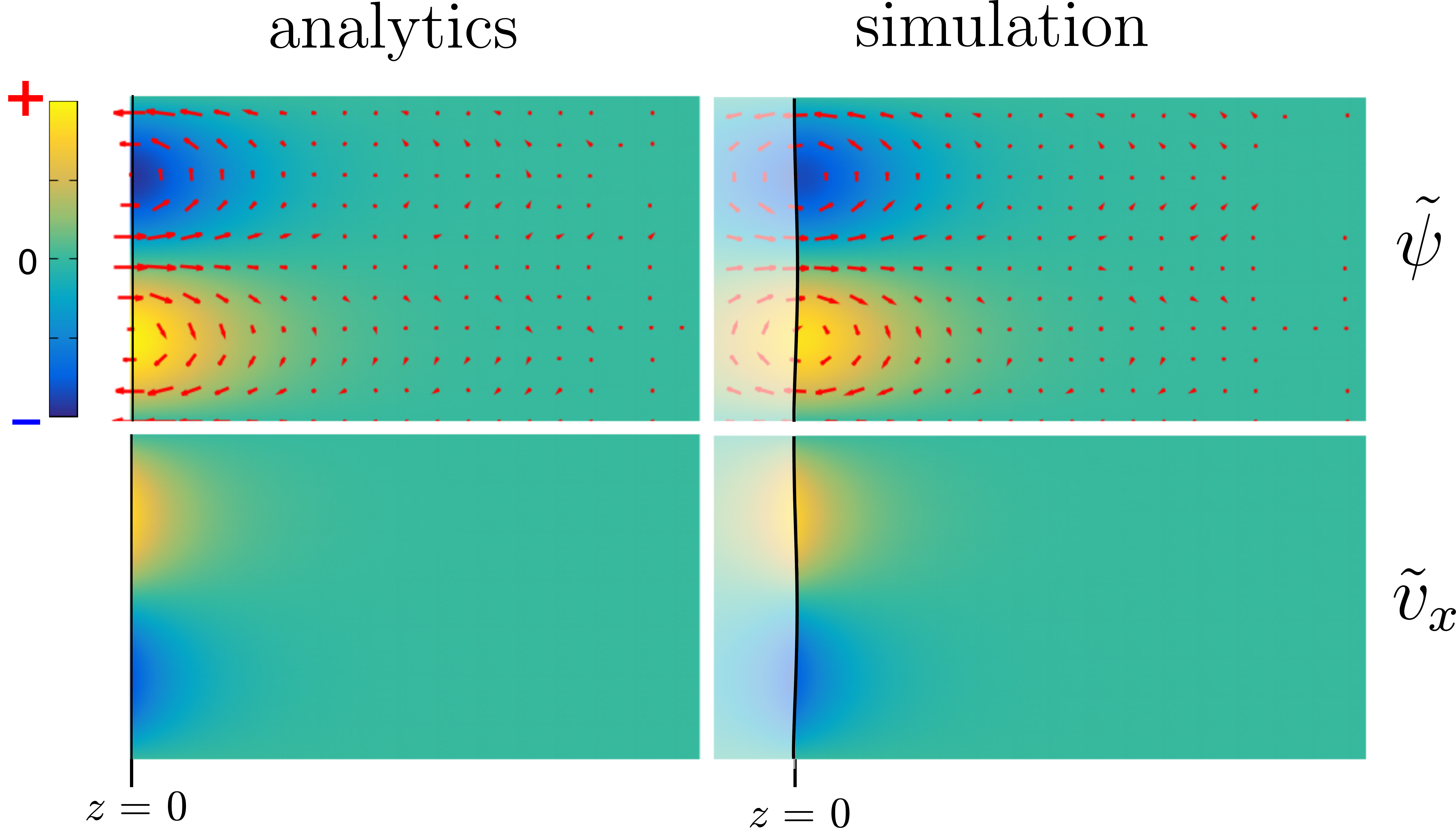}
  \caption{Top: colourmaps of the eigenfunction for the in-plane streamfunction, $\tilde{\psi}(y,z)$, with corresponding in-plane velocity vectors shown as red arrows. Bottom: colourmaps of the out-of-plane velocity component, $\tilde{v}_x$. For each quantity, the analytical solution of Eqn.~\ref{eqn:eigenfunction} is shown (left column) alongside the corresponding quantity extracted from the early time regime of our full nonlinear simulations (right column).  In each colourmap, the $y$ coordinate is vertical and the $z$ coordinate is horizontal, with the (unperturbed) interface at $z=0$. The region occupied by air in the simulation is shown translucent; no air is considered in the analytical calculation. Overall amplitude of colourscale is arbitrary. Results are for the Johnson-Segalman model in a biperiodic flow geometry. Slip parameter $a = 0.3$, surface tension $\Gamma / G L_y = 0$, imposed shear rate $\gdotb\tau = 0.125$. }
  \label{fig:eigenfns}
\end{figure}

\subsection{Criterion for edge fracture}

The linearised bulk equations, Eqns.~\ref{eqn:bulk_linear}, the linearised condition of interfacial force balance, Eqns.~\ref{eqn:interfacec} and the linearised equation of interfacial motion, Eqn.~\ref{eqn:moves}, can now together be solved to determine whether, for any given interfacial tension $\Gamma$ and imposed flow rate
$\gdot$, the heterogeneous perturbations grow towards an
edge fractured state, or decay to leave a flat interface.

\begin{figure}[!b]
  \includegraphics[width=\columnwidth]{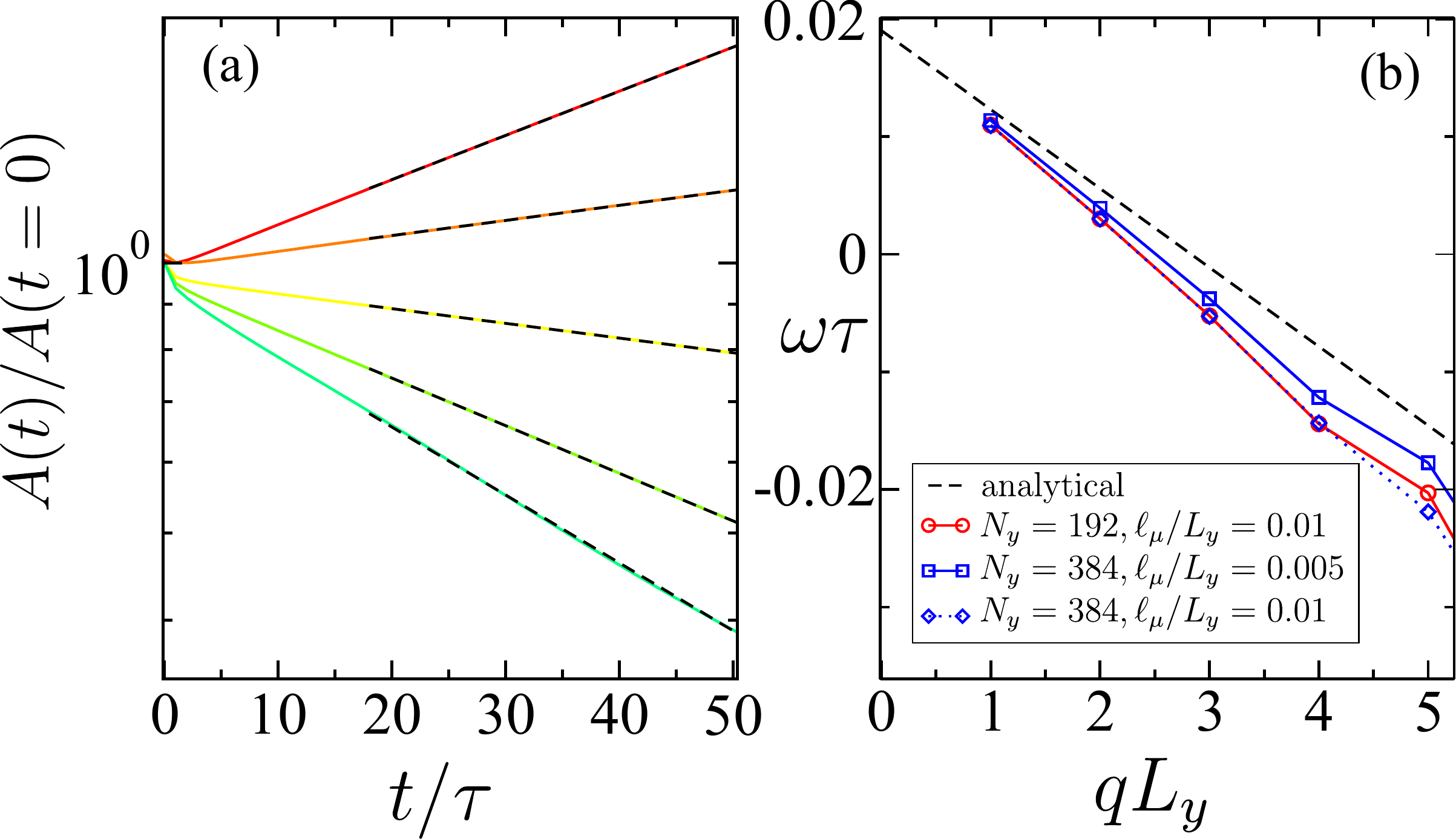}
  \caption{Left: time evolution $\exp(\omega(q)t)$ of the amplitudes of the modes $\exp(iqy)$ with the lowest five values of wavevector $q$ consistent with the boundary conditions. Best linear fits to these are shown as black dashed lines.  Right: resulting dispersion relation of growth rate as a function of wavevector, reconstructed from the slopes of these best fits, are shown by the symbols,  for three different numerical grids and values of the fluid-air interfacial thickness. Simulations are performed in the Johnson-Segalman model in the biperiodic flow geometry. Slip parameter $a=0.3$, interfacial tension $\Gamma/G L_y=0.0025$, imposed shear rate $\gdotb\tau = 0.25$, polymeric diffusion lengthscale $\lC = 0$. In the left panel, $N_y = 384$, $N_z =3840$ and $\lm/L_y = 0.005$.}
  \label{fig:dispersion}
\end{figure}

To leading order in
$\gdot$, and at any wavevector $q$ in the $y$ direction, the solution to the linearised bulk force balance condition, Eqn.~\ref{eqn:bulk_linear}, has the following normal mode form:
\bea
\tilde{\psi}(y,z,t)&=&\left[A e^{-qz}+B e^{-kz}\right]e^{iqy}e^{\omega t},\nonumber\\
\tilde{v}_x(y,z,t)&=&C e^{-qz}e^{iqy}e^{\omega t}.\nonumber\\
\label{eqn:eigenfunction}
\eea
In these equations,
$k=q/\sqrt{1+\beta}$, with $\beta=G^2\tau^4(1+\alpha-a)(a-\alpha)\gdot^2/(G\tau+\etas)^2$. 
The constants $A,B$ and $C$ can be determined by imposing the three boundary conditions of componentwise force balance across the fluid-air interface, Eqns.~\ref{eqn:interfacec},  although we do not write out the expressions for $A,B,C$ here.  

A colourmap of this analytical solution for the eigenfunction $\tilde{\psi}(y,z)$ and $\tilde{v}_x(y,z)$ at a fixed time $t$ is shown in the left two panels of Fig.~\ref{fig:eigenfns}. Excellent agreement is obtained with  the form extracted from the early time regime of our full nonlinear simulation, shown in the right two panels. As can be seen, the perturbation due to any interfacial disturbance of
wavelength $2\pi/q$ decays on the same lengthscale $O(q^{-1})$ into the
bulk.

The corresponding eigenvalue $\omega$ follows by inserting the normal mode solution for the interfacial position:
\be
\tilde{h}(y,t)=iq e^{iqy}e^{\omega t}
\ee
together with the solution of Eqn.~\ref{eqn:eigenfunction} for the streamfunction into the linearised equation of interface motion, Eqn.~\ref{eqn:moves}.  This gives
\be
\omega=\frac{1}{2\tau}\left[\frac{1}{2}\Delta\sigma\frac{d|N_2|}{d\gdot}\middle/\frac{d\sigma}{d\gdot}-\Gamma q\right],
\label{eqn:eigenvalue}
\ee
to within a small correction $O(\etas/G\tau)$.

\begin{figure}[!t]
  \centering
  \includegraphics[width=\columnwidth]{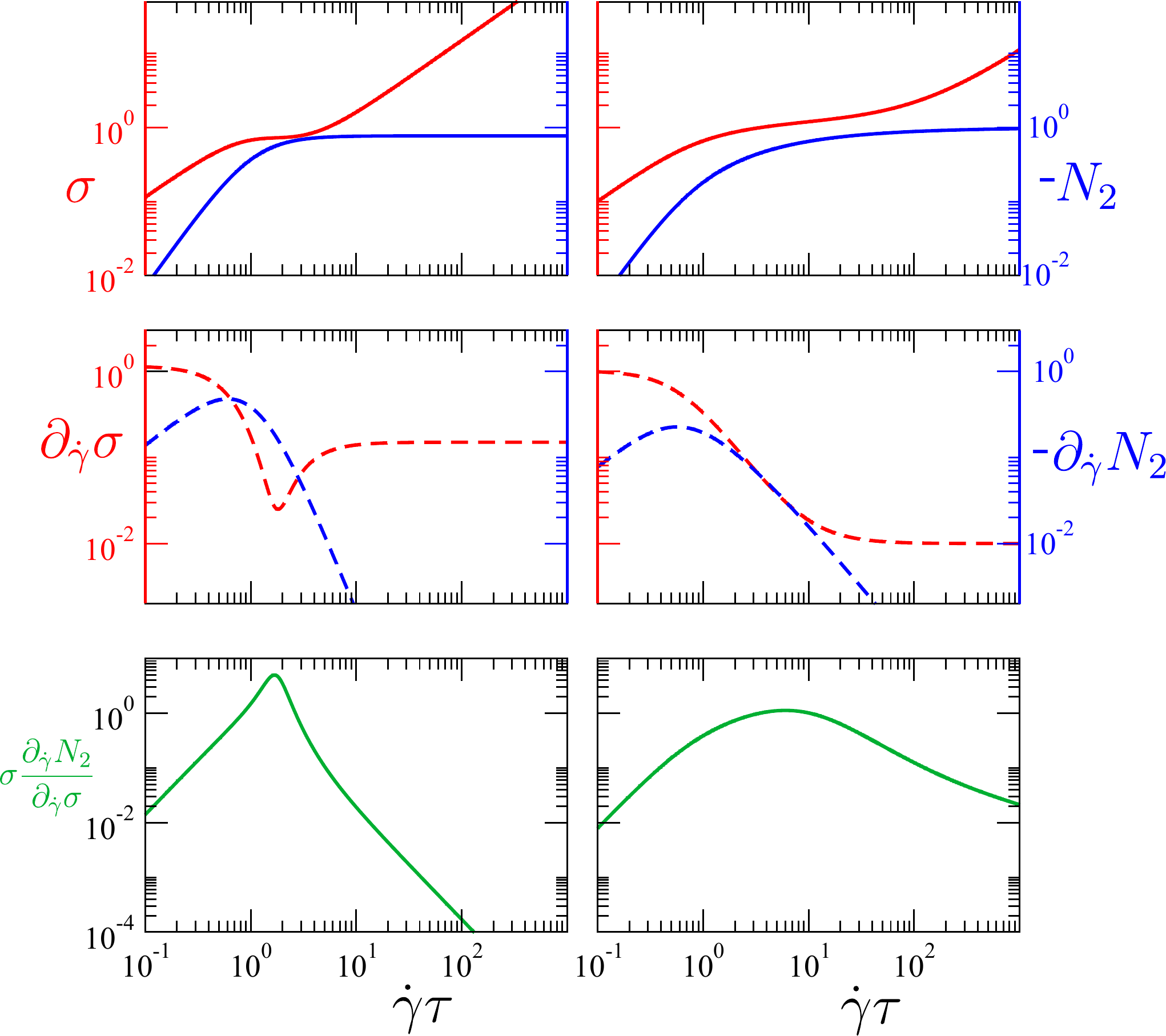}
  \caption{Rheological quantities appearing in the criterion for the onset of edge fracture in Eqn.~\ref{eqn:criterion}, plotted as a function of shear rate. Left: Johnson-Segalman model, slip parameter $a=0.3$. Right: Giesekus model $\alpha=0.4$.}
  \label{fig:contributions_both}
\end{figure}

In the left panel of Fig.~\ref{fig:dispersion}, we plot as coloured solid lines the time-evolution of the logarithm of the amplitude of the five modes with the lowest values of $q$ consistent with the boundary conditions, as obtained in the early time regime of our full nonlinear simulations. Each should follow the linear form $\omega(q)t$. Accordingly, we perform a best fits (black dashed lines) to the slopes of these lines. These best-fit slopes are plotted as a function of wavevector in the right panel to construct dispersion relation $\omega(q)$ of growth rate as a function of wavevector. We have plotted this as solid lines in the right panel, from simulations with three different numerical grids and fluid-air interfacial thicknesses.  The results of the analytical prediction of Eqn.~\ref{eqn:eigenvalue} (which assumes zero interfacial thickness) are shown by the black dashed line for comparison, with good agreement at low wavevectors.

Perturbations grow at any
wavevector $q$  if their eigenvalue $\omega(q)>0$.
This condition is most readily satisfied for the mode with
the lowest wavevector that is consistent with the biperiodic boundary conditions,
$q=2\pi/L_y$, consistent with the dispersion relation having its maximum value at the lowest $q$ in Fig.~\ref{fig:dispersion} (right). Accordingly, our final criterion for an initially flat
fluid-air interface to undergo edge fracture is given by
\be
\frac{1}{2}\Delta\sigma\frac{d|N_2(\gdot)|}{d\gdot}\bigg/\frac{d\sigma}{d\gdot}>\frac{2\pi\Gamma}{L_y}.
\label{eqn:criterion}
\ee
This criterion (after rescaling by the factor 2 discussed at the end of Sec.~\ref{sec:nonlinear}, to account for the difference between the biperiodic and walled geometries) is marked by the dashed line in Fig.~\ref{fig:PD}, and
gives good agreement at low shear rates with the onset of
fracture in our full nonlinear numerical simulations. An alternative, equivalent form of this criterion is to write, more simply:
\be
\frac{1}{2}\Delta\sigma \frac{d|N_2|}{d\sigma}>\frac{2\pi\Gamma}{L_y}.
\label{eqn:criterionb}
\ee

Appearing on the left side of this onset criterion, written in the form (\ref{eqn:criterion}), are several rheological quantities pertaining to the base state, as follows. First is the shear stress $\sigma(\gdot)$, which is shown as a function of shear rate by the solid red line in Fig.~\ref{fig:contributions_both} (top). (Recall that $\Delta\sigma=\sigma$, within our assumption that the outside air viscosity is zero.) Second, in the denominator, is the derivative of this quantity, $d\sigma/d\gdot$, which is shown by the red dashed line in Fig.~\ref{fig:contributions_both} (middle). This is the local slope of the flow curve, sometimes referred to as the tangent viscosity. Third is the derivative of the amplitude of the second normal stress with respect to strain rate, $\partial_{\gdot}|N_2(\gdot)|$, which is shown by the blue dashed line in Fig.~\ref{fig:contributions_both} (middle). These three quantities all combine to give the quantity plotted in green in Fig.~\ref{fig:contributions_both} (bottom), which indeed follows the shape of the instability thresholds in Fig.~\ref{fig:PD}. The region of re-entrant stability at high strain rates can therefore now be understood as arising from the decrease of $\partial_{\gdot}|N_2(\gdot)|$ at large strain rates, consistent with $|N_2(\gdot)|$ saturating to a constant as $\gdot\tau\gg 1$.

It is worth a reminder at this point that the criterion derived in this section applies to the biperiodic flow geometry. In order to apply to the experimentally realisable case of flow between hard walls, the box size must be rescaled by a factor $2$, as described at the end of Sec.~\ref{sec:nonlinear} and in the caption of Fig.~\ref{fig:robust}.

\begin{figure}[!t]
  \centering
  \includegraphics[width=\columnwidth]{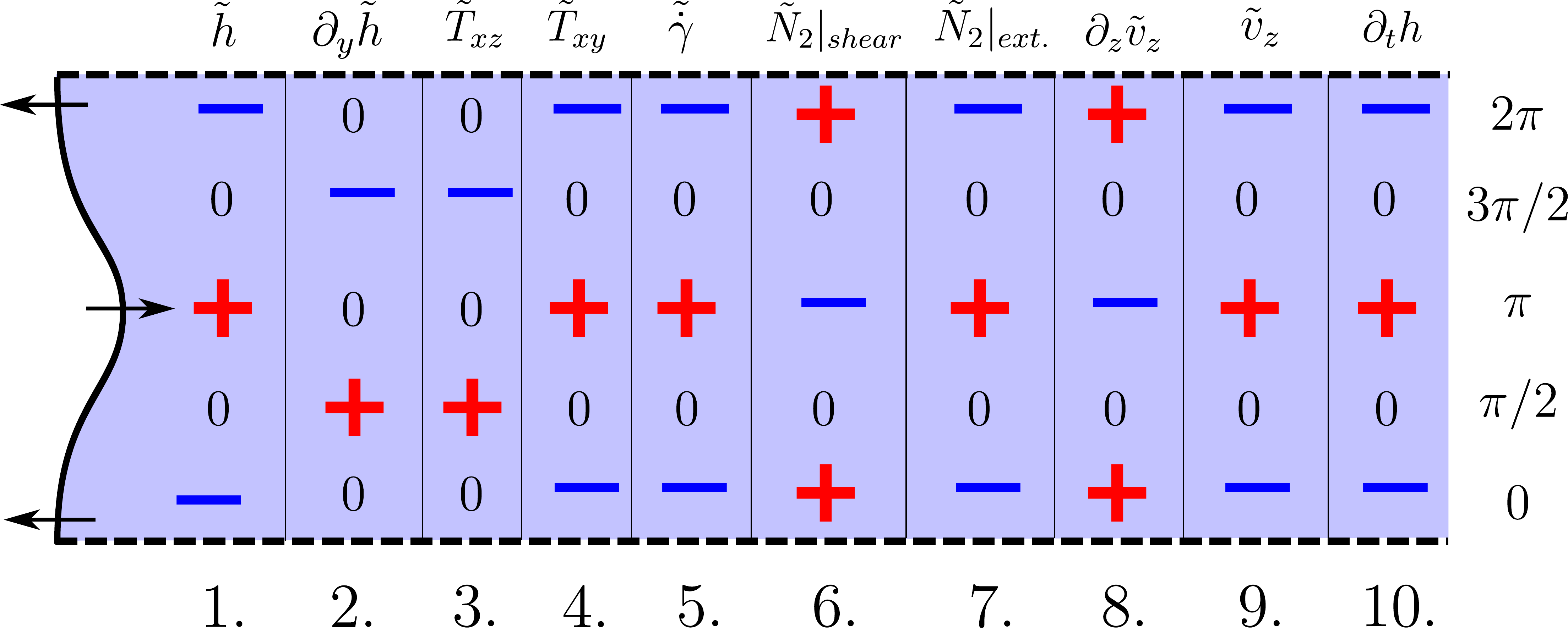}
  \caption{Schematic explaining the positive feedback mechanism in the edge fracture instability. The signs of the quantities denoted by the symbols at the top of the figure are shown at five different $y-$locations across the channel, given an interfacial disturbance of the form as sketched at the left.}
  \label{fig:mechanism}
\end{figure}

\begin{figure*}[!t]
  \includegraphics[width=\textwidth]{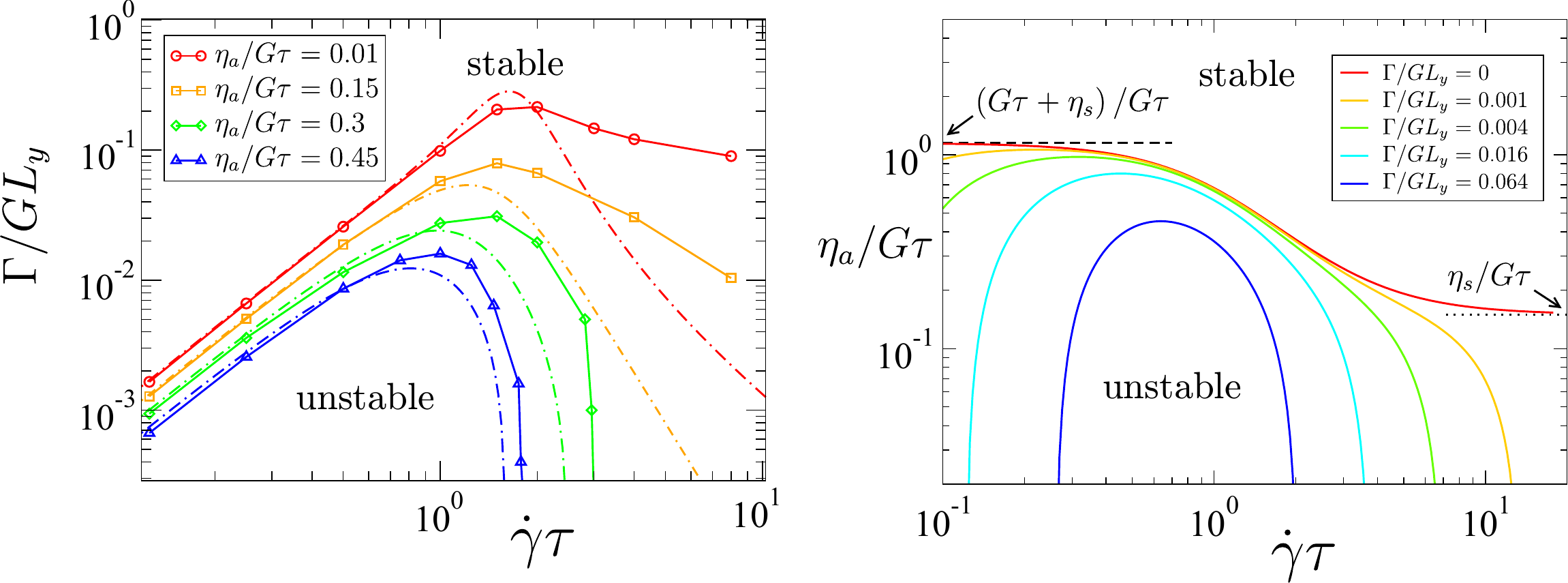}
  \caption{Threshold for onset of edge fracture in the Johnson-Segalman model in biperiodic shear. Left: thresholds shown in the plane of surface tension $\Gamma / G L_y$ and shear rate $\gdot\tau$, for several values of the viscosity $\etaa / G \tau$ of the bathing medium. Solid lines: full nonlinear simulation. Dot-dashed lines: linear stability analysis, valid in limit $\gdot\tau \to 0$. Right: in the plane of viscosity of the bathing medium  $\etaa / G \tau$ and shear rate $\gdot\tau$, for several values of the surface tension $\Gamma / G L_y$. Slip parameter:  $a = 0.3$.}
  \label{fig:mitigate}
\end{figure*}

\subsection{Mechanism of edge fracture}
\label{sec:mechanism}

The results of our analytical calculation allow us to understand the basic mechanism of the edge fracture instability, as follows.  Were the
interface between the fluid and air to remain perfectly flat, the jump $\Delta\sigma$ in shear
stress across it would be consistent with force balance.  Imagine now that a small interfacial tilt $\partial_y\tilde{h}$ arises, as shown in the second  column of
Fig.~\ref{fig:mechanism}. This  exposes the jump in shear stress across the interface, potentially disturbing the $x-$component of force balance across the interface, as expressed in the first of Eqns.~\ref{eqn:interfaceb}. To recover this $x-$component of interfacial
force balance, a counterbalancing perturbation
$\tilde{T}_{xz}|_{z=0^+}=iq h\Delta\sigma$ is required, as shown in the third column of Fig.~\ref{fig:mechanism}.  To maintain the $x$-component of force
balance in the fluid bulk (recall the first of Eqns.~\ref{eqn:bulk}), a corresponding
perturbation $\tilde{T}_{xy}$ is needed, as sketched in the fourth column of Fig.~\ref{fig:mechanism}. This is achieved via a
perturbation  $\tilde{\gdot}=\partial_y\tilde{v}_x=qh\Delta
\sigma/\sigma'(\gdot)$ in the shear rate (fifth column of
Fig.~\ref{fig:mechanism}).  The second normal stress $N_2
\approx -(1-a+\alpha)\gdot^2$
in the fluid bulk then suffers a corresponding perturbation 
$\tilde{T}_{yy}-\tilde{T}_{zz}|_{\rm shear}=-qh\Delta\sigma\,
|N_2|'(\gdot)/\sigma'(\gdot)$ (sixth column of
Fig.~\ref{fig:mechanism}). This must be counterbalanced  by an equal and opposite extensional
perturbation:
$\tilde{T}_{yy}-\tilde{T}_{zz}|_{\rm
ext}=4\partial_y\partial_z\tilde{\psi}=-4\partial_z\tilde{v}_z=4q\tilde{v}_z$ (seventh column of Fig.~\ref{fig:mechanism}). This in turn
demands a $z$-component of fluid velocity gradient (eight column of
Fig.~\ref{fig:mechanism}) and so of fluid velocity (ninth column). This finally further convects the interface,
$\partial\tilde{h}/\partial t=\tilde{v}_z=\tfrac{1}{4}\Delta\sigma
h\,|N_2|'(\gdot)/\sigma'(\gdot)$, enhancing the original interfacial tilt (tenth column of Fig.~\ref{fig:mechanism}) with a
growth rate
$\omega=\tfrac{1}{4}\Delta\sigma\,|N_2|'(\gdot)/\sigma'(\gdot)$.
This is indeed
consistent with Eqn.~\ref{eqn:eigenvalue} at zero surface tension (which is the limit in which the argument of this paragraph has been constructed). This mechanism resembles that of
instabilities between layered viscoelastic
fluids as studied (with different interfacial and wavevector orientations) in previous works~\cite{hinch1992instability,wilson1997short,nghe2010interfacially}.

\subsection{Comparison with Tanner's prediction}

We now compare our criterion for the onset of edge fracture, written again here for clarity:
\be
\frac{1}{2}\Delta\sigma\frac{d|N_2(\gdot)|}{d\gdot}\bigg/\frac{d\sigma}{d\gdot}>\frac{2\pi\Gamma}{L_y}.
\label{eqn:criterion1}
\ee
with Tanner's original prediction:
\be
 |N_2|>2\Gamma/3R.
 \ee
In Tanner's argument, $R$ is the radius of an (artificially)
assumed initially semicircular crack in the sample edge. To establish a closer correspondence between Tanner's criterion and ours, $R$
must now must be replaced by the dominant wavelength of instability
$L_y$.  To within $O(1)$ prefactors, the 
difference between Tanner's prediction and ours then lies in replacing
\be
|N_2|\to\frac{1}{2}\;\Delta\sigma\frac{d|N_2|}{d\gdot}\bigg/\frac{d\sigma}{d\gdot}.\;\;
\label{eqn:replace}
\ee
Given negligible air viscosity, the jump $\Delta\sigma$ in shear
stress across the interface between the fluid and air simply equals
the shear stress $\sigma$ in the fluid bulk, as noted above.  For most fluids, in the limit
of small shear rates, $N_2\sim -\gdot^2$ and
$\sigma=\eta\gdot$. Tanner's $|N_2|$ on the left hand side of of (\ref{eqn:replace})
then simply equals our expression on the right hand side. (In non-Brownian hard sphere suspensions, these scalings no longer hold. For example, the second normal stress scales linearly with the shear rate. It would be interesting in future work to study the edge fracture instability in that class of materials.) 

The identification by Tanner of the role of $N_2$ in driving edge fracture was a remarkable early insight, and indeed Tanner discussed carefully a mechanism for edge fracture based on radial stress balance. However, this fortuitous agreement between the expressions given above should not be over-interpreted. Indeed, whereas our expression can be connected directly to the detailed mechanism of instability by tracing through the argument in the previous subsection, no such detailed mechanistic insight can be gained from Tanner's prediction. Furthermore, Tanner's prediction says nothing about the role of the shear stress in driving the instability.

At higher
shear rates, the simple power law scalings $\sigma\sim\gdot$ and $N_2\sim-\gdot^2$ noted above no longer hold (in general), and
our prediction departs from Tanner's.  Indeed (assuming a monotonic $N_2(\gdot)$) Tanner predicts a monotonic phase boundary $\Gamma(\gdot)$ even for a material in which the derivative of $N_2$ with respect to shear rate is a non-monotonic function of shear rate. This is in stark contrast to our non-monotonic one, with its re-entrant stability at high strain rates.

\section{Possible mitigation strategy}
\label{sec:mitigation}

Finally, and perhaps most importantly, the results of our linear stability calculation suggest a recipe
via which edge fracture might be mitigated in experimental practice. We recall, in particular, that the left hand side of our onset criterion (\ref{eqn:criterion}) contains the term $\Delta\sigma$, which is the jump in shear stress between the fluid and outside medium. In any experiment where the outside medium is air, the shear stress outside the fluid will be negligible and we simply have $\Delta\sigma=\sigma$.

Instead immersing the flow cell
in an immiscible Newtonian `bathing fluid' with a viscosity larger
than that of air, more closely matched to that of the study-fluid, the
jump $\Delta\sigma$ in shear stress between the study and bathing
fluids will clearly be
reduced.  We explore this suggestion in  Fig.~\ref{fig:mitigate}a), showing the thresholds for onset of instability for successively increasing values of $\etaa$, indeed finding greater stability at larger $\etaa$. 
The dot-dashed lines in Fig.~\ref{fig:mitigate} show the results of a linear stability analysis performed as in Sec.~\ref{sec:linear} above, but generalised to include a non-zero viscosity of the outside air. Fig.~\ref{fig:mitigate}b) explores the suppression of instability for different levels of surface tension, as a function of $\etaa$ and shear rate.

Another obvious strategy for mitigating edge fracture would be to try ensure as large an interfacial surface tension as possible, by suitable choice of the (Newtonian) bathing medium.

\section{Conclusions}
\label{sec:conclusion}

In this work, we have performed a detailed theoretical study of edge
fracture, combining direct nonlinear simulations with linear stability
analysis for the initial onset of edge fracture, and finding full
agreement between these. We have derived a criterion of the onset of
edge fracture, and provided a detailed understanding of the physical
mechanism that drives the instability. Our results also suggest a new
strategy via which edge fracture might potentially be mitigated in
experimental practice.  We also suggest that, in containing the second
normal stress difference, our criterion for the onset of edge fracture
may be used as a means to determine that quantity experimentally

Despite these successes, our work contains several notable limitations that should be addressed in future studies. We discuss these finally in turn.

First, we have considered only constitutive models that predict a negative second normal stress. Positive values of the $N_2$ were discussed as being potentially stabilising against edge fracture in Ref.~\cite{keentok1999edge}. It would be interesting in future work to re-do the present calculation in a constitutive model that predicts a positive $N_2$.

Second, we have considered only fluids with a finite terminal relaxation time, $\tau$, for which the shear stress scales as $\sigma\sim\gdot\tau$ and the second normal stress as $N_2\sim -(\gdot\tau)^2$ in the limit $\gdot\tau\ll 1$. It would be interesting in future work to consider the case of non-Brownian suspensions~\cite{denn2014rheology}, in which $N_2$ instead scales linearly with shear rate.

Third, we have assumed in the derivation of our criterion a base flow corresponding to a state of steady shear. Edge fracture is also widely seen in transient rheological tests such as shear startup. Future theoretical studies could profitably consider the effects of a time-dependent base state on the onset of edge fracture. 

Fourth, we have considered the limit of planar Couette flow, arguing that this geometry provides a good approximation to cylindrical Couette flow in the small gap limit, and to cone-and-plate (or plate-plate) flows for small gap and large device radius.  If this assumption is correct, our results should apply equally to cylindrical Couette and cone-and-plate. In view of this, an outstanding puzzle is why edge fracture is discussed much more commonly in the literature for cone-and-plate (or plate-plate) than for cylindrical Couette. One possibility is that the highly viscoelastic materials for which edge fracture occurs can only be studied in cone-and-plate (or plate-plate). Another possibility is that there is an additional   feature, not captured in our analysis, that renders cone-and-plate (or plate-plate) much more highly susceptible to edge fracture than cylindrical-Couette. For example, do the hoop stresses present in cone-and-plate (or plate-plate) greatly accelerate the nonlinear dynamics of edge fracture, even while leaving the criterion for initial onset unaltered? Indeed, in plate-plate and cone-plate devices, the crack propagates
inwards towards regions with smaller radius of curvature, whereas in cylindrical Couette it propagates orthogonally to the direction in which the curvature changes. A simpler explanation is that for the same size of crack, the fraction of the surface disturbed in curved Couette is smaller than in cone-plate (or plate-plate) due to the larger surface of the measuring fixture.

Fifth, we have ignored bulk instabilities that can lead to complicated secondary flows within the sample. We have thereby assumed that such instabilities are deferred until beyond the onset of edge fracture, and/or that edge fracture and bulk instabilities occur in different classes of materials. It would be interesting in future work to carry out a simulation study that allows a crossover, as a function of some material property, from a situation in which bulk instabilities first arise as a function of increasing strain rate, to one in which edge fracture dominates instead.

Sixth, and related, we have ignored the well known bulk instability in which an initially homogeneous shear flow gives way to the formation of coexisting shear bands. In recent works, it has been demonstrated theoretically that shear bands must lead inevitably to full edge fracture in some parameter regimes~\cite{skorski2011loss}; and conversely that even modest edge disturbances can lead to an apparent bulk shear banding, even in materials that would not show any true shear banding in the absence of edge effects~\cite{hemingway2018edge}. 
Indeed, possible
edge fracture in experiments concerning  shear banding in entangled polymers has
been discussed in Refs.~\cite{sui2007instability,li2015startup,schweizer2008departure,li2013flow,wang2014letter,li2014response,boukany2015shear}. Clearly, more remains to be done to understand the interplay between bulk shear banding and surface instabilities such as edge fracture, in both steady and transient flow protocols.

Seventh, we have ignored wall slip, which occurs widely in strongly sheared entangled polymers. Indeed, wall slip remains under studied from a theoretical viewpoint, with a notable paucity of constitutive models for the layer of fluid immediately adjacent to the wall of the flow cell, as compared with those for the bulk fluid. Future theoretical studies should consider the relative dominance of and/or interplay between edge fracture and wall slip.

Finally, our calculations have ignored inertia, working throughout in the creeping flow limit of zero Reynolds number. Inertia may be an importance factor in the non-linear dynamics of edge fracture, and should be considered in future simulation studies.

{\it Acknowledgements -- } The research leading to these results has
received funding from the European Research Council under the EU's 7th
Framework Programme (FP7/2007-2013) / ERC grant number 279365. We thank Mike Cates, Gareth McKinley, Peter Olmsted, Roger Tanner and Dimitris Vlassopoulos for enjoyable discussions.

\bibliography{EdgeFracture}
\bibliographystyle{apsrev4-1}

\end{document}